\documentclass[twocolumn,showpacs,pra]{revtex4-1}
\usepackage{amsmath}
\usepackage{amssymb}
\usepackage{graphicx}
\usepackage[unicode]{hyperref}
\hypersetup{
   unicode=true,          % non-Latin characters in AcrobatÕs bookmarks
   a4paper=true,
   plainpages=false,
   pdftitle={Title of PDF},    % title
   pdfauthor={Author of PDF},     % author
   pdfsubject={Subject of PDF},   % subject of the document
   colorlinks=true,       % false: boxed links; true: colored links
   citecolor=blue,        % color of links to bibliography
}
\usepackage[caption=false,position=top,singlelinecheck=off,justification=raggedright]{subfig}
\begin{document}

\title{Relaxation dynamics of ultracold bosons in a double-well potential: Thermalization and prethermalization in a nearly integrable model}

\author{Jayson G. Cosme}

\affiliation{New Zealand Institute for Advanced Study, Dodd-Walls Centre for Photonics and Quantum Technology, Centre for Theoretical Chemistry and Physics, Massey University Auckland, Private Bag 102904, North Shore, Auckland 0745, New Zealand}


\date{\today}
\begin{abstract}
We numerically investigate the relaxation dynamics in an isolated quantum system of interacting bosons trapped in a double-well potential after an integrability breaking quench. Using the statistics of the spectrum, we identify the postquench Hamiltonian as nonchaotic and close to integrability over a wide range of interaction parameters. We demonstrate that the system exhibits thermalization in the context of the eigenstate thermalization hypothesis (ETH). We also explore the possibility of an initial state to delocalize with respect to the eigenstates of the postquench Hamiltonian even for energies away from the middle of the spectrum. We observe distinct regimes of equilibration process depending on the initial energy.  For low energies, the system rapidly relaxes in a single step to a thermal state. As the energy increases towards the middle of the spectrum, the relaxation dynamics exhibits prethermalization and the lifetime of the metastable states grows.
Time evolution of the occupation numbers and the von Neumann entropy in the mode-partitioned system underpins the analyses of the relaxation dynamics.
\end{abstract}
\maketitle

\section{Introduction }

Motivated by experiments in ultracold atoms \cite{Kinoshita2006, Bloch2008},
there has been a resurgence of interest in exploring the
fundamental aspects of statistical mechanics and whether statistical properties can emerge from unitary time evolution.
To this end, recent attempts in describing thermalization
in closed quantum systems focused on time-evolving 
observables that are local to a subsystem and how they approach thermal equilibrium described by Gibbs ensemble \cite{Polkovnikov2011,Rigol2008}.
One proposed mechanism to understand thermalization is the eigenstate thermalization hypothesis (ETH) \cite{Deutsch1991,Srednicki1994,Rigol2008}.

The ETH conjectures that under certain conditions, the expectation value of a physically relevant observable will have a long-time average close to an appropriate microcanonical ensemble prediction.
When testing the ETH, the system is usually driven out of equilibrium by a quantum quench in the parameters of the Hamiltonian. 
The ETH was numerically tested against other hypotheses 
and it was shown that for initial states following the conditions of the  ETH, thermalization is achieved
within observables \cite{Rigol2008,Rigol2012}.
To introduce basic ideas of the ETH, consider an initial
state $|\phi_{0}\rangle$, which is not an eigenstate of some postquench or final
Hamiltonian $\hat{H}$ without any degeneracies. Then, the initial state can be expanded in terms of the eigenstates
of $\hat{H}$ with eigenvalues $E_{k}$ as  
\begin{equation}
|\phi_{0}\rangle=\sum_{k}\alpha_{k}|k\rangle.
\end{equation}
After the quench, the initial state will undergo time evolution as 
\begin{equation}
|\phi(t)\rangle=e^{-i\hat{H}t/\hbar}|\phi_{0}\rangle
\end{equation}
and therefore, the expectation value of an
observable is given by
\begin{equation} \label{etheq} \langle\hat{\mathcal{O}}(t)\rangle=\sum_{k,l}\alpha_{l}^{\ast}\alpha_{k}e^{i(E_{l}-E_{k})t/\hbar}\mathcal{O}_{lk}
\end{equation}
with $\mathcal{O}_{lk}=\langle l|\hat{\mathcal{O}}|k\rangle$. Due to dephasing, Eq.~\eqref{etheq} will relax to the infinite-time average, 
\begin{equation}\label{ethex}
\overline{\langle\hat{\mathcal{O}}\rangle}=\lim_{t \to \infty}\frac{1}{t}\int_0^{t}d\tau\langle \hat{\mathcal{O}}(\tau) \rangle=\sum_{k}|\alpha_{k}|^{2}\mathcal{O}_{kk}.
\end{equation}
This long-time average is commonly referred to as the diagonal ensemble prediction due to the diagonal ensemble, $\hat{\rho}_{d}=\sum_{k}|\alpha_{k}|^{2}|k\rangle\langle k|$.
According to the ETH, the diagonal ensemble average of an observable, Eq.~\eqref{etheq}, is close to an appropriate microcanonical average $\langle \hat{\mathcal{O}} \rangle_{\mathrm{ME}}$,
\begin{align}\label{ethme}
&\langle \hat{\mathcal{O}} \rangle_{\mathrm{ME}} = \mathcal{N}^{-1}\sum_{|E_k-E_0|<\Delta E} \mathcal{O}_{kk}, \\ \nonumber
&\overline{\langle\hat{\mathcal{O}}\rangle}=\sum_{k}|\alpha_{k}|^{2}\mathcal{O}_{kk}, \\ \nonumber
&\langle \hat{\mathcal{O}} \rangle_{\mathrm{ME}} = \overline{\langle\hat{\mathcal{O}}\rangle},
\end{align}
where $\mathcal{N}$ is the number of eigenstates within a narrow energy window $\Delta E$ around the initial mean energy of the system $E_0$.
The last equality in Eq.~\eqref{ethme} is true if the eigenstate expectation value (EEV), $\mathcal{O}_{kk}$, is a smooth
function of $E_{k}$, while the off-diagonal elements ${\cal O}_{kl}$
are negligible
\cite{Srednicki1994,Rigol2008}. There is no rigorous analytical proof of the ETH but different numerical studies on spin systems and lattice models both for small and large system sizes
support its validity \cite{Rigol2009, Santos2010c, Genway2012, Steinigeweg2014, Sorg2014, Rigol2014, Kim2014, Khodja2015, Alba2015}.

Another important aspect of relaxation is the question of how local observables equilibrate. This is particularly important in the dynamics of nearly integrable systems, which may exhibit a two-stage equilibration process. Both theoretical and experimental studies have investigated the nature of prethermalization in various physical setups \cite{Moeckel2008,Kollar2011,Gring2012,Marcuzzi2013,Smith2013,Gong2013,Essler2014,Nessi2014,Landea2015,Bertini2015,Langen2015}. 

The emergence of statistical relaxation can also be linked to the onset of quantum chaos as highlighted by previous studies \cite{Flambaum1997,Santos2012,Santos2012c}. In this context, particular emphasis is laid on the form of the energy distribution of an initial state and how it compares with predictions of quantum chaos. We utilize this approach to demonstrate chaotic initial states even for energies far from the middle of the spectrum of eigenenergies, an extension of the findings in Ref.~\cite{Torres2013}.

In this work, we study the nonequilibrium dynamics of interacting bosons in a one-dimensional double-well potential, which can be modelled using a two-site multilevel Bose-Hubbard Hamiltonian \cite{Garcia2012,Fialko2012}. 
Contrary to previous studies dealing with few atoms in multiple lattice sites, we study the other limit of having finite large number of bosons confined in few sites. As a matter of fact, it was checked that this system will thermalize due to the ergodicity of the associated mean-field trajectories in the semiclassical limit of large $N$ \cite{Cosme2014}.
Here, we only consider $N=35$ bosons and use exact diagonalization to access the eigenvalues and the eigenstates of the final Hamiltonian. First, we determine the chaoticity of the postquench Hamiltonian on the basis of spectral statistics. We then verify the validity of the ETH in this system. Our results based on the distribution of EEV illustrate that not all eigenstates are thermal. 
Using all possible initial product state configurations, we compare the expectation values of local operators obtained using the diagonal and the microcanonical ensemble. 

Most of the studies related to subsystem thermalization involve spatial partitioning of the system \cite{Santos2012b, Sorg2014,Khlebnikov2014,Zhang2015}. Instead, we implement a different partitioning scheme where a subsystem will consist of only one mode, either the lower or the upper level in one of the wells. Since thermalization is a dynamical process, we also  compute the time evolution of local operators and the von Neumann entropy of a subsystem. 
We find that the relaxation dynamics of the von Neumann entropy is instructive in revealing prethermalization in the system. Moreover, we observe that fast relaxation of a local operator correlates with a rapid single-step growth of the von Neumann entropy as it approaches the Gibbs entropy. On the other hand, we observe a two-stage relaxation process for initial states close to the middle of the spectrum. Finally, we discuss a possible mechanism for the prethermalization in the system and the associated time scales.

This paper is organized as follows. In Sec.~\ref{sec:modquench}, we first describe the model and the quench protocol used in this work. Section \ref{sec:levelspac} contains results that characterize the spectrum and the chaoticity of the postquench Hamiltonian.
We investigate the requirements of the ETH in Sec. \ref{sec:eth}.
The key results on relaxation and thermalization in the system are discussed in Sec. \ref{sec:quenchdyn}. 
In Sec.~\ref{sec:deme}, we compare the long-time and microcanonical averages of local operators. The presence of chaotic initial states across the spectrum is discussed in Sec. \ref{sec:chaos}.
Results for the actual time evolution of local properties in the system are presented in Sec \ref{sec:relmod} and \ref{sec:sub}.
In Sec. \ref{sec:pretherm}, we discuss the relaxation time scales and the process of prethermalization in the system. Finally, we briefly summarize our findings in Sec. \ref{sec:conc}.

\section{Model and Quench Dynamics}\label{sec:modquench}

We study thermalization in a finite isolated quantum system with bosons trapped in a double-well potential
shown in Fig. \ref{fig:Model}. A single boson can occupy four modes (two single-particle levels in each well), which can be described by the following
two-level generalization of the Bose-Hubbard model \cite{Garcia2012,Fialko2012}: 
\begin{figure}[!htb]
 \includegraphics[width=0.7\columnwidth]{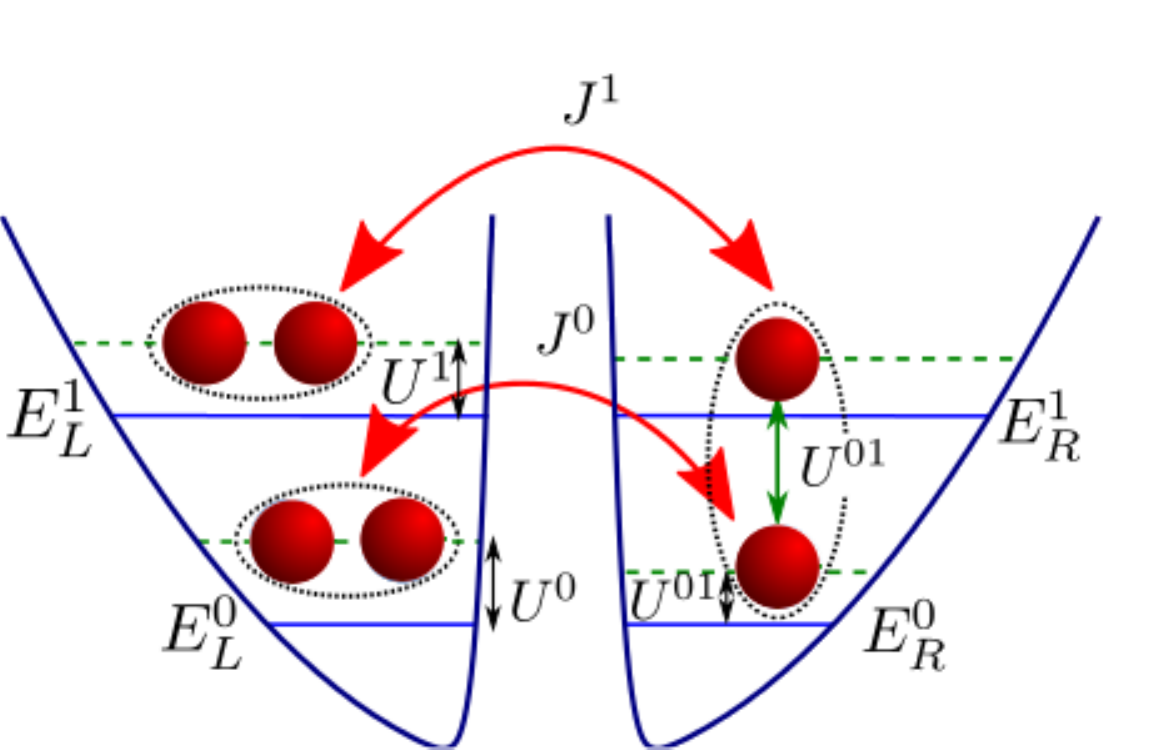} \protect\caption{(Color online) Schematic of the double-well potential with two energy levels. The interlevel coupling
$U^{01}$ adds another degree of freedom in the system, which is expected to break integrability. \label{fig:Model}}
\end{figure}
\begin{align}\label{eq:Hamilt}
\hat{H} & =-\sum_{r\neq r',\ell}J^{\ell}\hat{b}_{r}^{\ell\dagger}\hat{b}_{r'}^{\ell}+\sum_{r,l}U^{\ell}\hat{n}_{r}^{\ell}(\hat{n}_{r}^{\ell}-1)+\sum_{r,\ell}E_{r}^{\ell}\hat{n}_{r}^{\ell}\nonumber \\
 & +U^{01}\sum_{r,\ell \neq \ell'}(2\hat{n}_{r}^{\ell}\hat{n}_{r}^{\ell'}+\hat{b}_{r}^{\ell\dagger}\hat{b}_{r}^{\ell\dagger}\hat{b}_{r}^{\ell'}\hat{b}_{r}^{\ell'}),
\end{align}
where $\hat{b}_{r}^{\ell\dagger}$ and $\hat{b}_{r}^{\ell}$ are the bosonic creation and annihilation operators of an atom in well $r$ and energy level $\ell$, respectively. The Hamiltonian in Eq. \eqref{eq:Hamilt} can be derived
from its second-quantized form after a tight-binding approximation \cite{Garcia2012}.
The system is effectively mapped into a two-site lattice model with added dimensionality or
degree of freedom due to the coupling with the first excited single-particle level.

We consider an external harmonic potential with oscillator frequency $\omega_{0}$, which
is split by a focused laser beam located at the center of the trap and described by a Gaussian potential $V_{0}\mathrm{exp}(-x^{2}/2\sigma^{2})$.
The parameters in Eq. \eqref{eq:Hamilt} can
be easily evaluated for a specific realization of the double-well potential described by the single-particle Hamiltonian \cite{Garcia2012},
\begin{equation}\label{hsp}
\hat{H}_{\mathrm{sp}} = \frac{-\hbar^2}{2m}\frac{\partial^2}{\partial x^2} + \frac{m\omega^2x^2}{2} + V_0\mathrm{exp}\biggl(-\frac{x^2}{2\sigma^2}\biggr).
\end{equation}
The localized functions $\phi_{r}^{\ell}$ in level $\ell \in \{0,1\}$ and site $r \in \{L,R\}$
are obtained by symmetric and antisymmetric superpositions of the eigenstates of the single particle Hamiltonian in Eq. \eqref{hsp}. 
The corresponding eigenvalues are 
$E_{r}^{\ell}=\int dx\phi_{r}^{\ell*}(x)\hat{H}_{\mathrm{sp}}\phi_{r}^{\ell}(x)$.
The tunneling terms between wells are $J^{\ell}=-\int dx\phi_{L}^{\ell*}(x)\hat{H}_{\mathrm{sp}}\phi_{R}^{\ell}(x)$.
The interaction terms between atoms in the same well and the same energy level are $U^{\ell}=g\int dx|\phi_{r}^{\ell}|^{4}$. There is also an interlevel coupling corresponding to the two-atom hopping term between energy levels in the same well given by $U^{01}=g\int dx|\phi_{r}^{0}(x)|^{2}|\phi_{r}^{1}(x)|^{2}$. 
The limit of $U^{01}=0$ is the integrable
Bose-Hubbard dimer model for a double well, which was extensively studied before in different contexts \cite{Aubry1996, Smerzi1997, Milburn1997, Albiez2005, Levy2007, Gillet2014}.
Integrability of the dimer can be broken by adding one or more degrees of freedom without changing the number of conserved quantities, which for the dimer model corresponds to the energy and the number of bosons \cite{Hennig1996,Flach1997}. 
A finite value of $U^{01}$ introduces additional degrees of freedom and thus breaks the integrability of the system. 
Note that upon choosing the trap parameters, the only free parameter in the system is the coupling constant $g$, which affects both $U^{\ell}$ and $U^{01}$. This is a subtle detail that we need to keep in mind as blindly increasing the interaction $g$ (in hopes of increasing the integrability breaking term $U^{01}$) will actually push the system towards a strongly interacting regime dominated by quasidegeneracies \cite{Garcia2012}.

We now discuss the quench protocol used in our study of the nonequilibrium dynamics in the system. 
We consider a partitioning of the system such that each mode is considered as a subsystem and the remaining three modes will act as an environment.
Initially, the number of bosons in each mode is fixed; i.e., the tunneling and the interlevel terms are set to zero,
such that the system starts as a product state of eigenstates of each subsystem. In particular, we choose a set of initial states corresponding to all possible Fock state configurations $|n\rangle = |n_0\rangle = |n^0_L, n^0_R , n^1_L, n^1_R\rangle$. 
The eigenstates $|k\rangle$ of the postquench Hamiltonian $\hat{H}$ with eigenvalues
$E_k$ can be expressed in terms of the Fock basis, 
\begin{equation}
 |k\rangle = \sum_n C^k_n |n\rangle.
\end{equation}
This expansion is similar to how the mean-field or unperturbed (in our case uncoupled) 
basis is physically motivated 
in studies related to chaotic eigenstates and its connection to thermalization of
isolated systems \cite{Flambaum1997}. 

We do a sudden quench by turning on the tunneling and the interlevel coefficients. 
The wavefunction unitarily evolves in time according to 
\begin{equation}\label{phitk}
 |\psi(t)\rangle = e^{-i\hat{H}t/\hbar}|n_0\rangle = \sum_k C^{k*}_{n_0} e^{-iE_k t/\hbar}|k\rangle.
\end{equation}
For brevity, we refer to the mean energy, which is conserved during the dynamics,
\begin{equation}
 E_0 = \langle n_0|\hat{H}|n_0 \rangle = \sum_k E_k |C^k_{n_0}|^2,
\end{equation}
as simply the initial energy of the system.

For the postquench Hamiltonian, the barrier height is chosen to be $V_{0}=5\hbar\omega_{0}$ with
width $\sigma=0.1l_{ho}$, where the harmonic oscillator length is
$l_{ho}=\sqrt{\hbar/m\omega_{0}}$.
The interaction coupling $g$ is chosen such that $NU^0/\hbar\omega_0 = \mathrm{const}$.
For this parameter space of the trap, 
the coefficients in the Hamiltonian are $J^{0}/\hbar\omega_{0}=0.26$, $J^{1}/\hbar\omega_{0}=0.34$, $E_{r}^{0}/\hbar\omega_{0}=1.25$, $E_{r}^{1}/\hbar\omega_{0}=3.17$,
$NU^{0}/\hbar\omega_0=\mathrm{const.}$, $U^{1}=3U^{0}/4$, and $U^{01}=U^{0}/2$. 

We use exact diagonalization to obtain the eigenvalues and the eigenvectors expressed in the Fock basis.
To this end, we obtain the $D \times D$ Hamiltonian matrix in the Fock basis where the size of the Hilbert space is $D=(N+3)!/(N!)(3!)=8436$.
Alternatively, the time evolution of the wave function can be calculated from the time-evolving expansion coefficients,
\begin{equation}
|\psi(t)\rangle = \sum_{\{n\}} c_n(t)|n\rangle  
\end{equation}
where $|n\rangle$ spans all possible Fock state configurations of the system. Temporal evolution of the expansion coefficients is computed using standard numerical integrator, e.g., seventh- or eighth-order Runge-Kutta method.

\section{Ratio of Consecutive Level Spacings Distribution}\label{sec:levelspac}

An interesting question for finite quantum systems is the relevance of chaoticity of the Hamiltonian in the study of thermalization \cite{Torres2013}. 
It was conjectured using a system of one-dimensional lattice with spin-$1/2$, that the occurrence of thermalization depends only on the level of delocalization of initial states \cite{Torres2013}. Furthermore, this condition has been shown to be sufficient for initial states close to the middle of the spectrum irrespective of integrability (or chaoticity) in both prequench and postquench Hamiltonian. 

The level statistics of a chaotic Hamiltonian was found to possess similar spectral properties as random matrices \cite{Brody1981,Bohigas1984}. One common measure of level statistics is the level spacing distribution, $s_k = E_{k+1}-E_{k}$, where $\{E_k\}$ is the set of eigenenergies in ascending order \cite{Brody1981}.
In a Hamiltonian with time-reversal symmetry, the chaotic regime is identified by a level spacing distribution similar to the Gaussian orthogonal ensemble (GOE) \cite{Brody1981}. The connection between nonintegrability and a GOE-like spectral behavior was pointed out in various works such as Refs.~\cite{Montambaux1993,Kollath2010,Santos2010b,Santos2012c}.
On the other hand, integrable models are expected to have a Poissonian level spacing distribution as demonstrated in Refs.~\cite{Berry1977,Kollath2010,Santos2010b,Santos2012c}.

Due to the large interaction strengths involved in our quench dynamics, i.e., $NU^{\ell}/J^{\ell} > 1$
for both levels ${\ell}$, it is convenient
to avoid any unfolding procedure of the spectrum \cite{Kollath2010}. Instead, we obtain the distribution of the ratio of consecutive gaps between adjacent levels, 
$r_k = \mathrm{min}(s_k,s_{k-1})/\mathrm{max}(s_k,s_{k-1}) = \mathrm{min}(\frac{s_k}{s_{k-1}},\frac{s_{k-1}}{s_{k}})$ \cite{Oganesyan2007, Kollath2010, Atas2013}.
We compare the numerically obtained distribution $P(r)$ with the analytical expressions for the Poisson and the GOE distributions \cite{Atas2013}
\begin{equation}
P_\mathrm{P}(r)=\frac{2}{(1+r)^2},~P_\mathrm{GOE}(r)=\frac{27}{4}\frac{r+r^2}{(1+r+r^2)^{5/2}}.
\end{equation}
A more quantitative comparison can be made using the mean value of $r$ denoted as $\langle r \rangle$,
which is $\langle r \rangle_\mathrm{P} = 2\mathrm{ln}2 - 1 \approx 0.3863$ for the Poisson distribution and
$\langle r \rangle_\mathrm{GOE} \approx 0.5359$ for the GOE \cite{Atas2013}.

\begin{figure}[!ht]
\begin{center}
 \hspace*{0.5cm}\includegraphics[width=0.7\columnwidth]{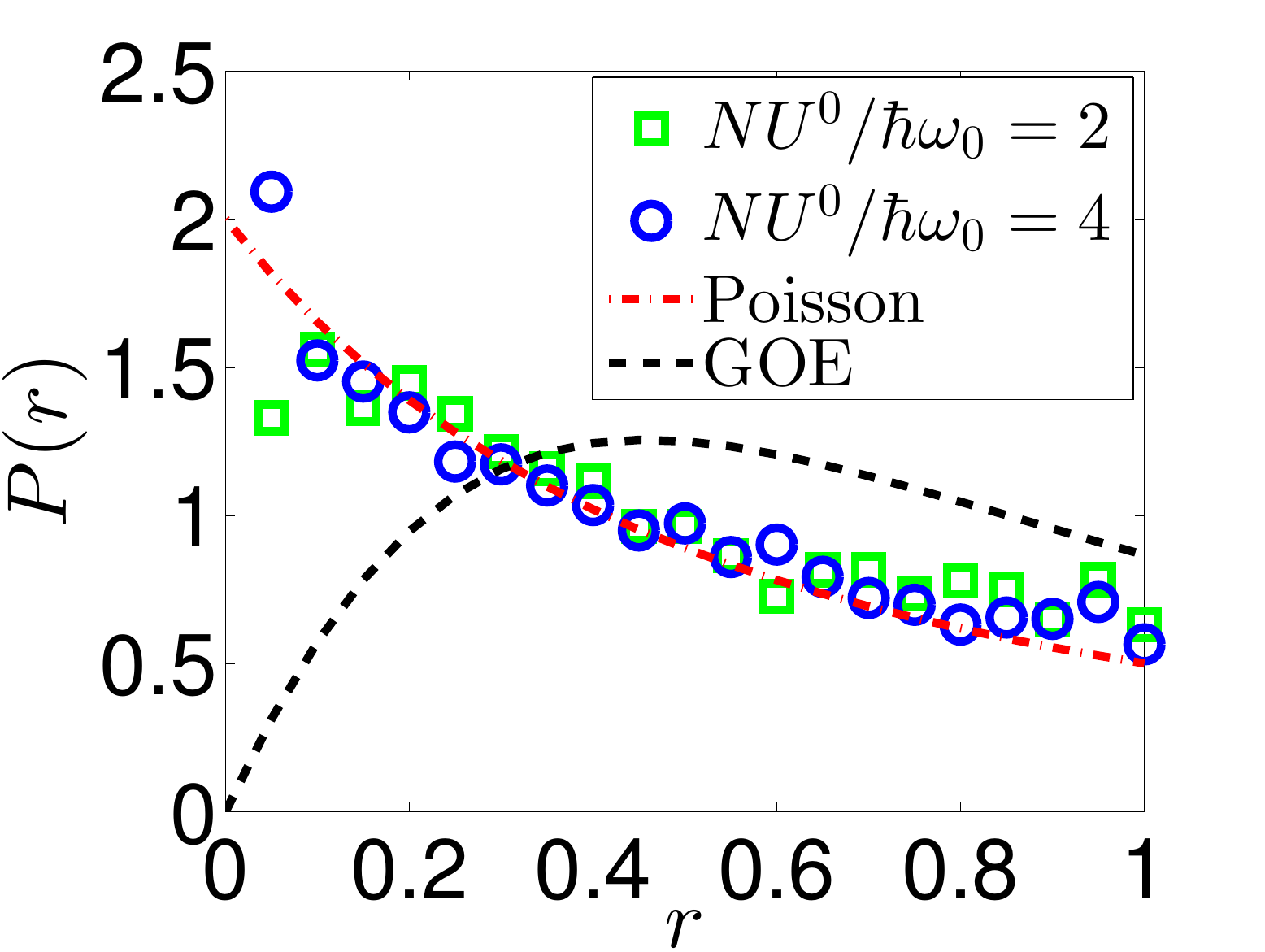}
\par\bigskip
\includegraphics[width=0.65\columnwidth]{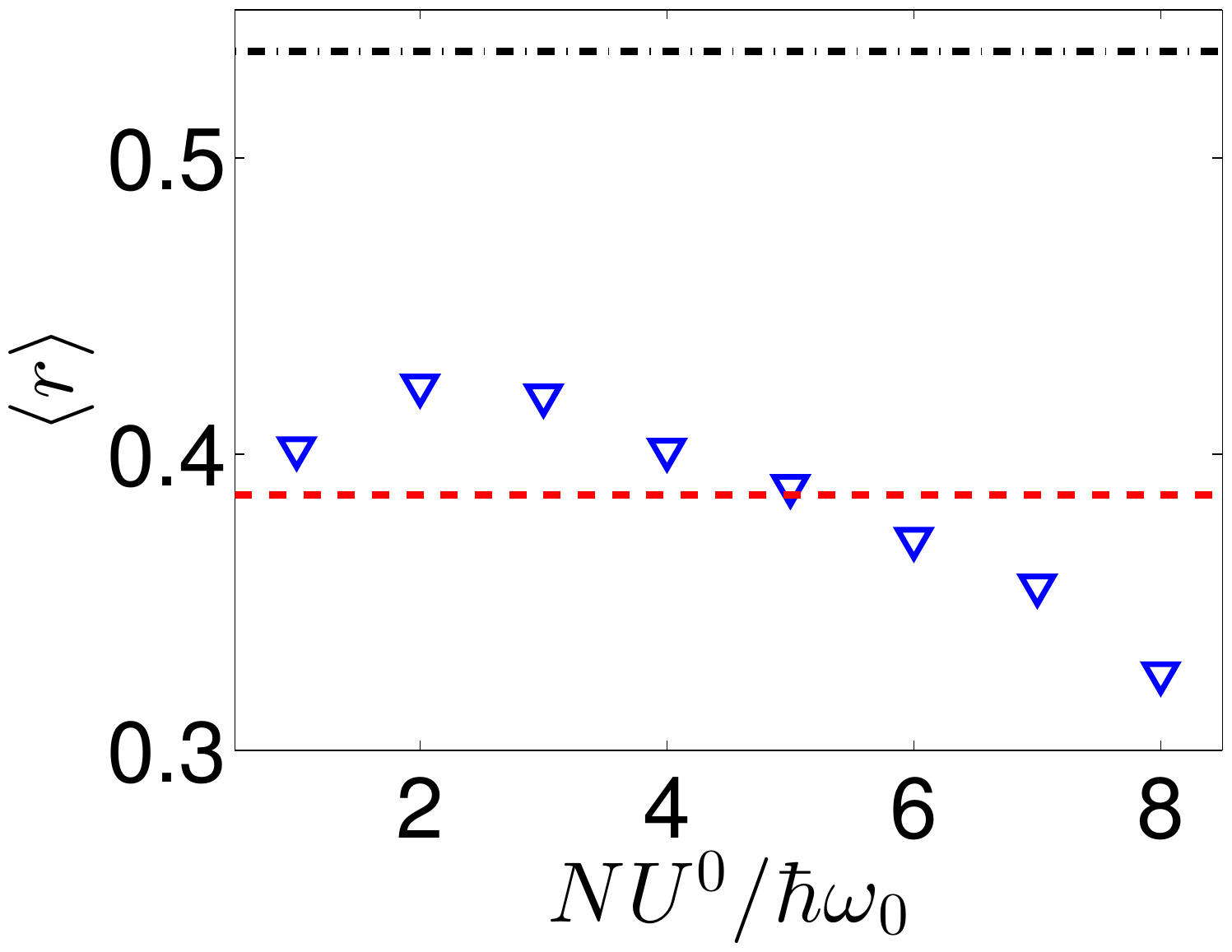}
\protect\caption{(Color online) (Top) Distribution of the ratio of adjacent level spacings $r$. (Bottom) Mean value $\langle r \rangle$ as a function of the interaction parameter.
The GOE (dashed-dotted line) and the Poissonian (dashed line) averages are also shown. }
\label{lsr}
\end{center}
\end{figure}

There is parity or reflection symmetry about the center of the double-well system.
In order to avoid mixing eigenenergies from different symmetry sector,
we separate the eigenenergies according to its parity, i.e., even or odd eigenvectors in the Fock basis. We do this by constructing eigenstates with well-defined parity in Fock space \cite{Mumford2014}. For the analysis of spectral statistics presented in this section, we numerically diagonalize the Hamiltonian in the well-defined parity basis.
We denote the reflection operator, which exchanges the Fock state from left to right or vice versa as $\hat{\mathcal{P}}|n^0_L,n^0_R,n^1_L,n^1_R\rangle = |n^0_R,n^0_L,n^1_R,n^1_L\rangle$. Then, we use a set of basis states given by:
\begin{equation}
|\widetilde{n}\rangle_{\pm} = \frac{1}{\sqrt{2}}\biggl(|n\rangle \pm \hat{\mathcal{P}}|n\rangle\biggr).
\end{equation}

The distribution of the ratio of adjacent energy gaps $P(r)$ for the even-parity sector is illustrated in Fig. \ref{lsr}.
First, we analyze the spectral statistics of the full spectrum.
Notice how $\langle r \rangle$ increases with the interaction strength before reaching its maximum around $NU^0/\hbar\omega_0=2$. It then decreases as you further increase the interaction strength. Perhaps more interesting, the mean value of $r$  becomes smaller than the Poisson distribution $\langle r \rangle_\mathrm{P}$ for $NU^0/\hbar\omega_0>5$. This level clustering can be explained by the appearance of quasidegenerate eigenvectors (in each symmetry sector) when the interaction terms $U^{\ell}$ become sufficiently larger than the tunneling terms $J^{\ell}$ \cite{Garcia2012}. Specifically, these quasidegenerate pairs appear as high-lying excited states of the system. 
Nevertheless, our results for the distribution of the ratio of consecutive level gaps imply that the postquench Hamiltonian is nonchaotic in the chosen parameter space of the trap. Even though the final Hamiltonian is nonintegrable, the spectral statistics of the final Hamiltonian is actually closer to the Poisson distribution than the GOE for the interaction strengths considered here. This indicates that the system is still close to an integrable point after the quench. 

In the following section, we demonstrate that thermalization is still viable, in the context of the ETH, for the nonchaotic Hamiltonian considered here. We also argue later that the presence of quasidegeneracies in the spectrum has important implications for the process of relaxation in the system.

\section{Eigenstate Thermalization Hypothesis}\label{sec:eth}

The microcanonical ensemble can describe the long-time averages if the requirements of the ETH are satisfied.
Specifically, the diagonal ensemble prediction for a local operator $\overline{\langle \hat{A} \rangle}$ will be close to a value predicted by a thermal or microcanonical ensemble if the EEV is a smooth function of $E_k$. This requirement is fulfilled if the vertical width of the EEV at a chosen energy is small \cite{Sorg2014}. We now calculate the eigenstate expectation values of the mode occupation number, $\langle k | \hat{n}^{\ell}_r | k \rangle$, and check how they are distributed in the eigenenergies of the system for different interaction strengths. 
\begin{figure}[!ht]
\subfloat[]{\includegraphics[width=0.5\columnwidth]{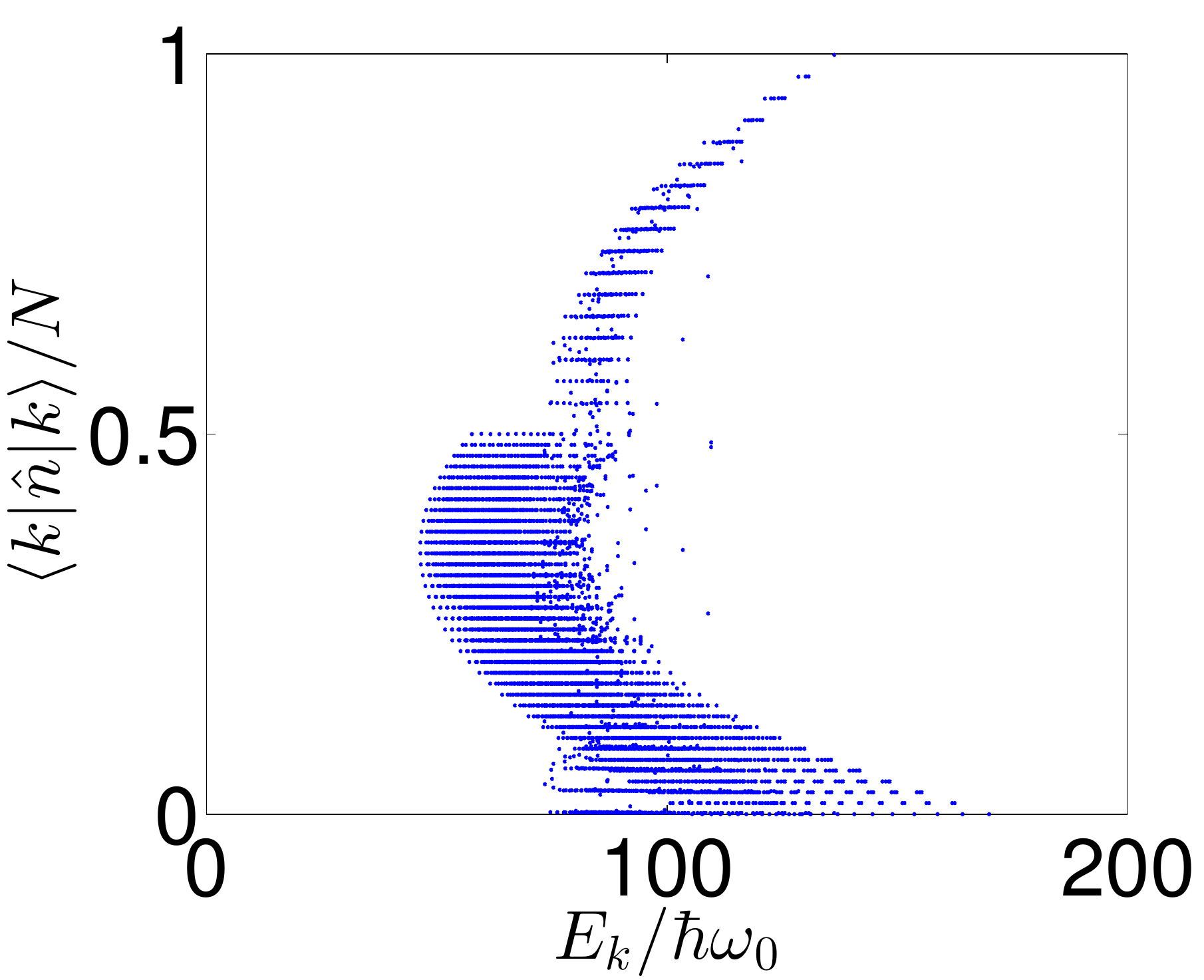}}\subfloat[]{\includegraphics[width=0.5\columnwidth]{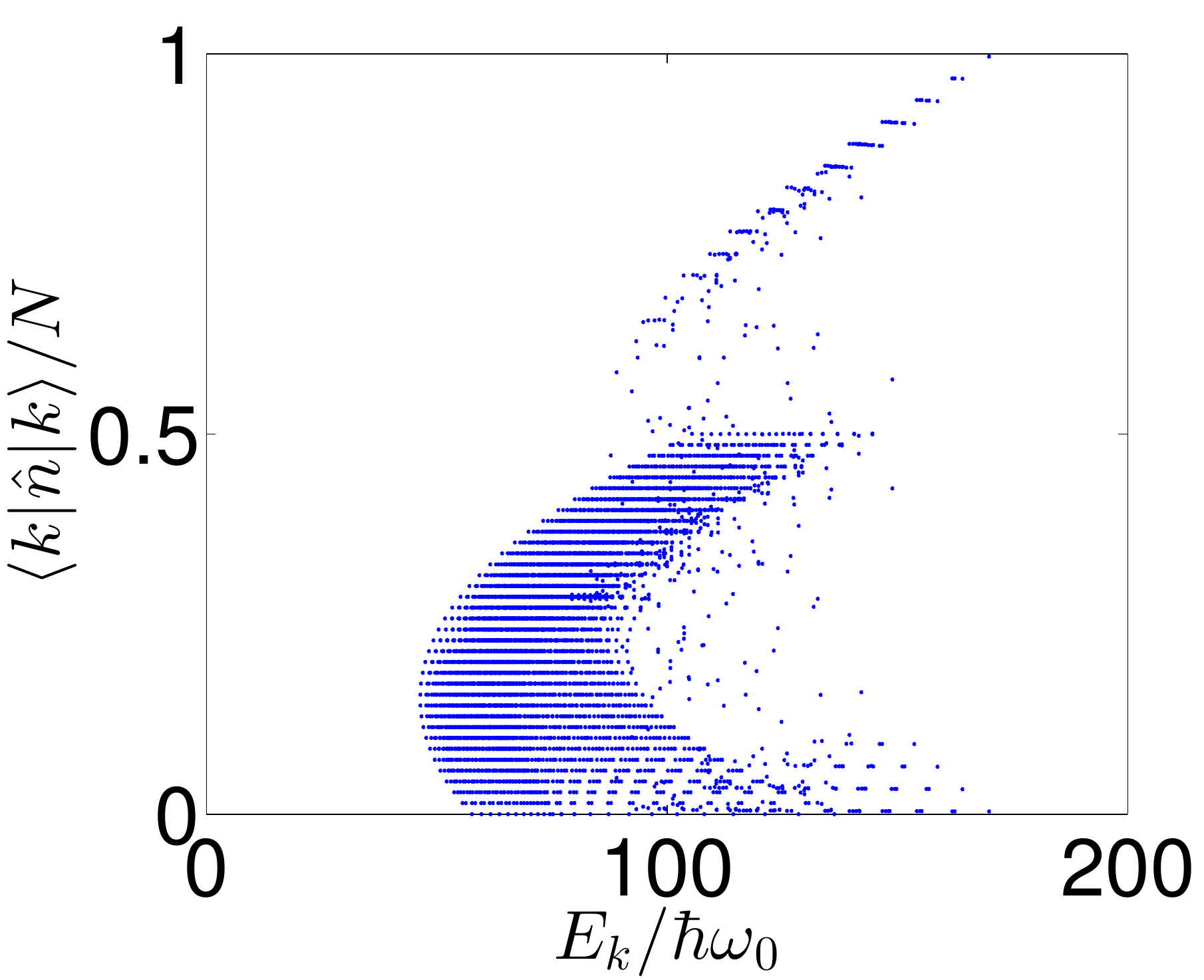}}
\protect\caption{(Color online) Distributions of EEV for $NU^{0}/\hbar\omega_0=4$ but $U^{01}=0$. (a) $\hat{n}^0_L$ (b) $\hat{n}^1_L$. }
\label{eev1} 
\end{figure}

First, we consider the integrable limit of uncoupled Bose-Hubbard dimers when the interlevel coupling is absent, $U^{01}=0$.  We plot in Fig. \ref{eev1}(a) the EEV of the dimer with the lower on-site energy while Fig. \ref{eev1}(b) shows the EEV of the dimer with higher on-site energy. The overall structure of the distribution of EEV appears similar in both dimers. That is, the expectation values of the mode occupation seemingly appear separated in flat bands, which then prohibits the EEV from being a smooth function of the energy. Physically, this is a direct consequence of $NU^{\ell}/J^{\ell} \gg 1$. There is large range of possible EEV at a given energy $E_k$ and hence thermalization must be absent in this limit.
\begin{figure}[!ht]
\subfloat[]{\includegraphics[width=0.5\columnwidth]{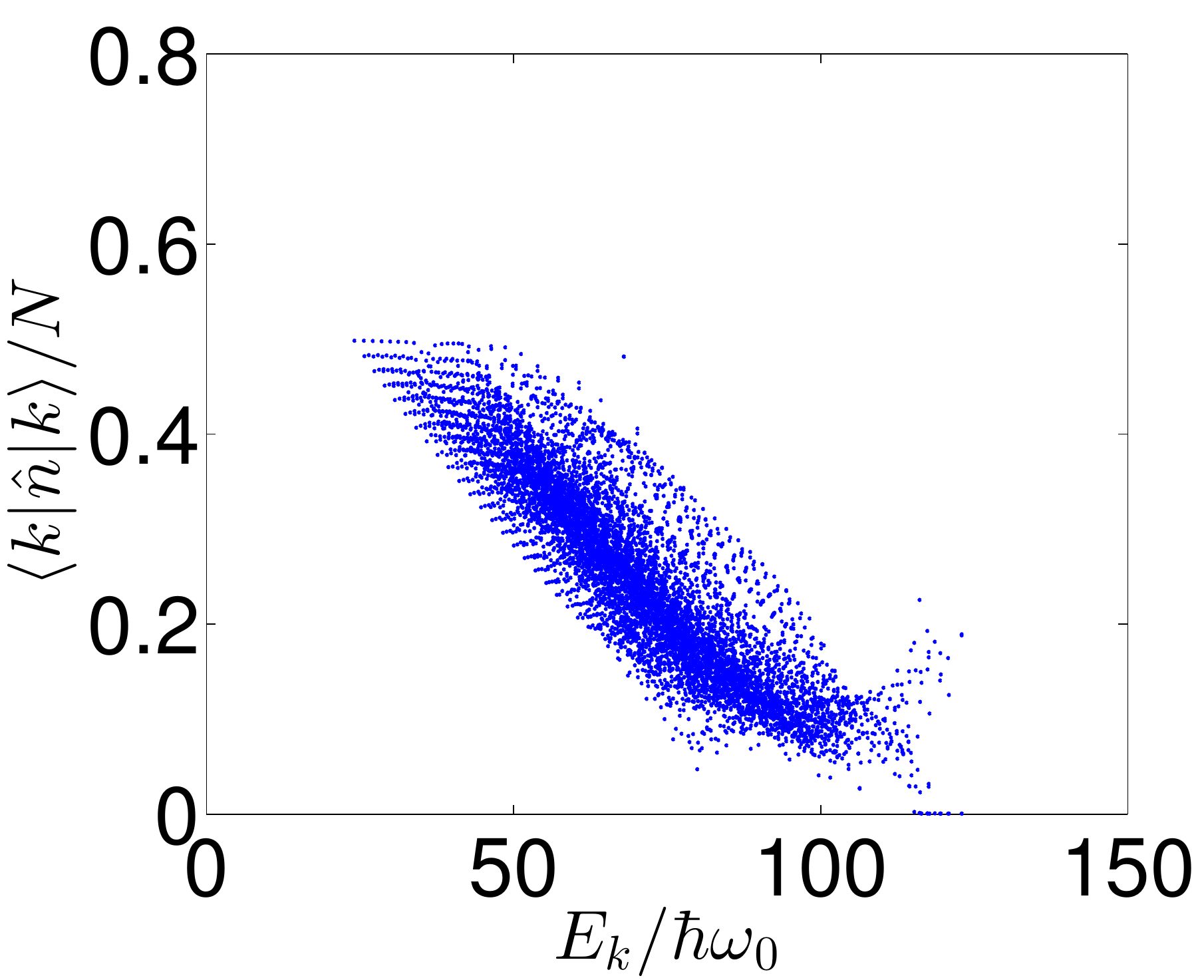}}\subfloat[]{\includegraphics[width=0.5\columnwidth]{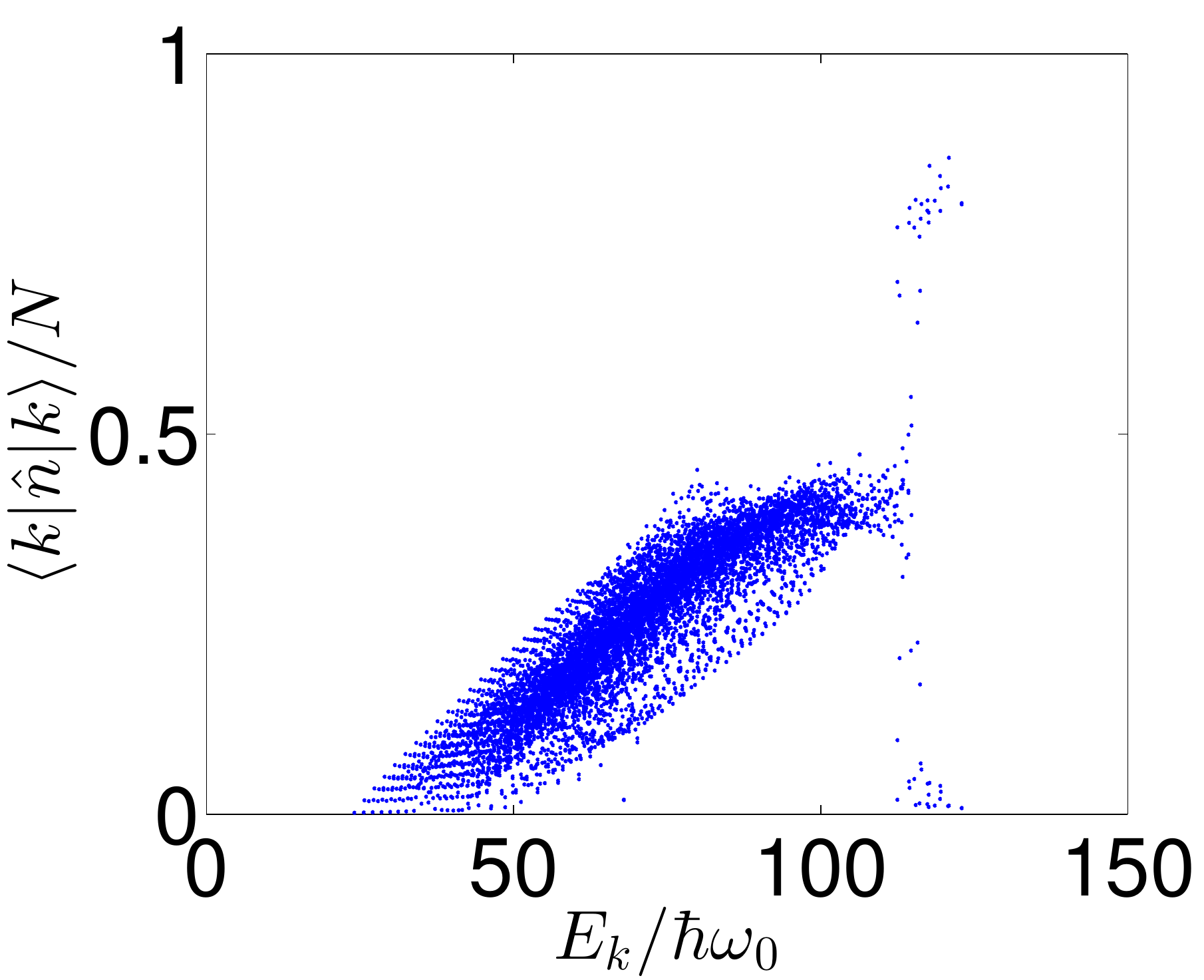}}\\
\subfloat[]{\includegraphics[width=0.5\columnwidth]{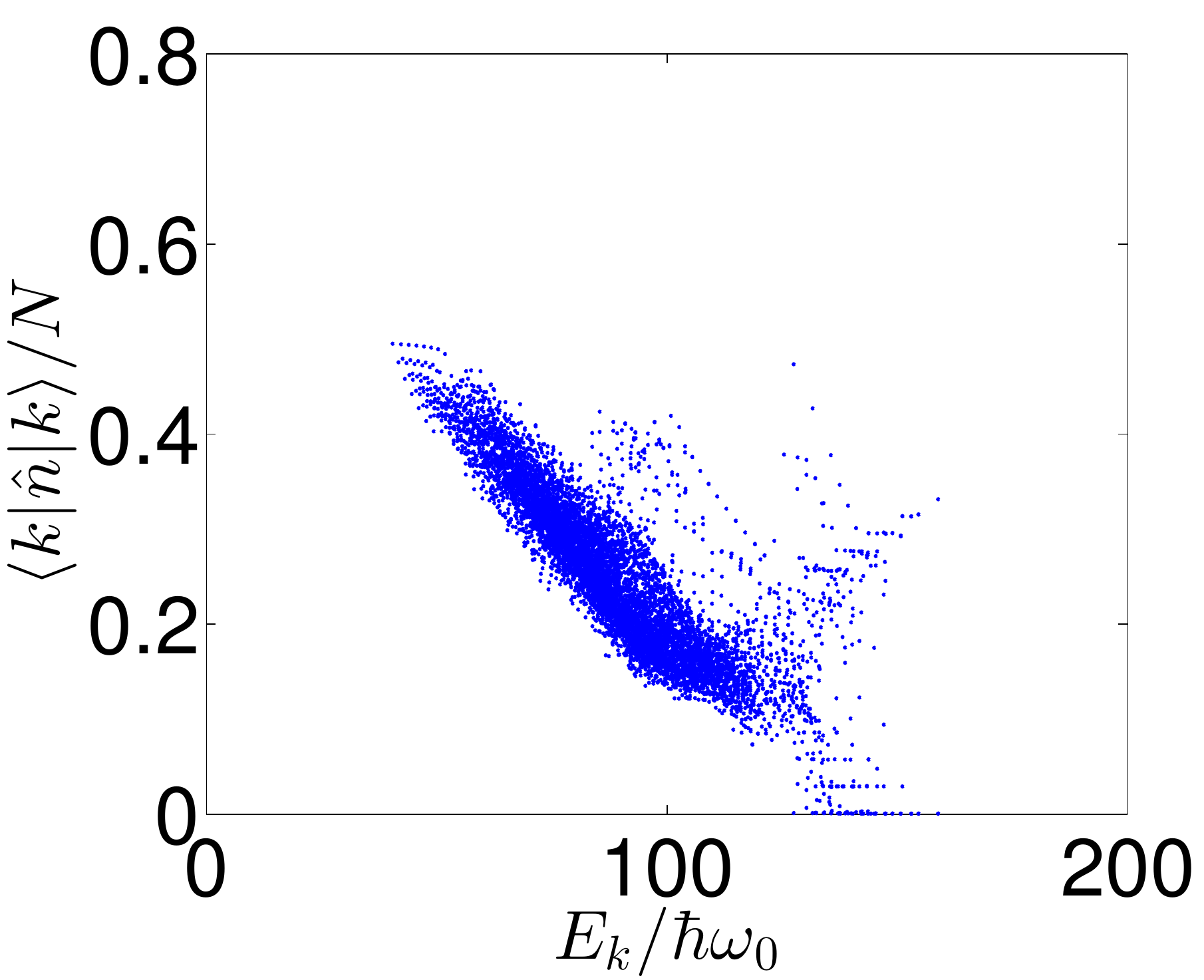}}\subfloat[]{\includegraphics[width=0.5\columnwidth]{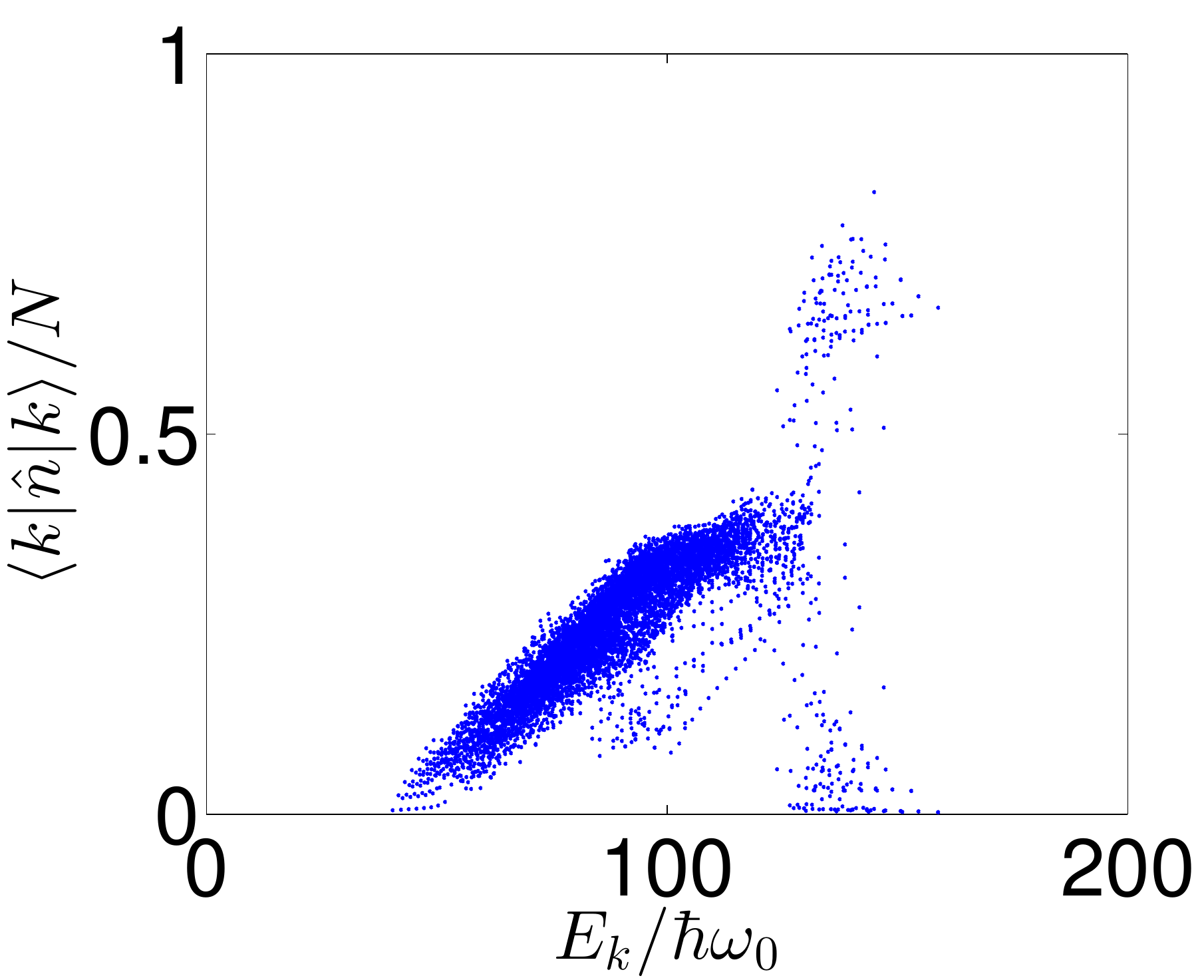}}\\
\subfloat[]{\includegraphics[width=0.5\columnwidth]{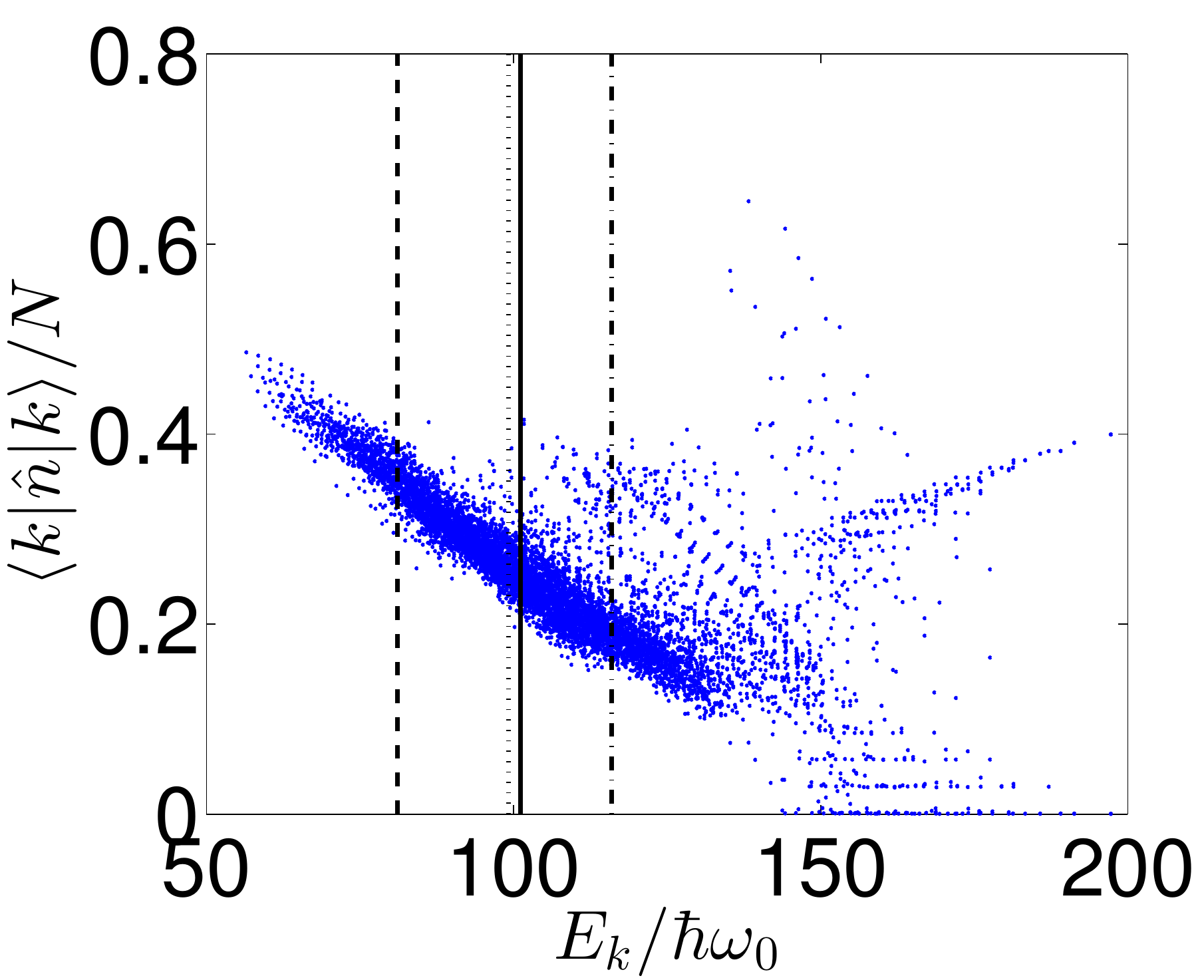}}\subfloat[]{\includegraphics[width=0.5\columnwidth]{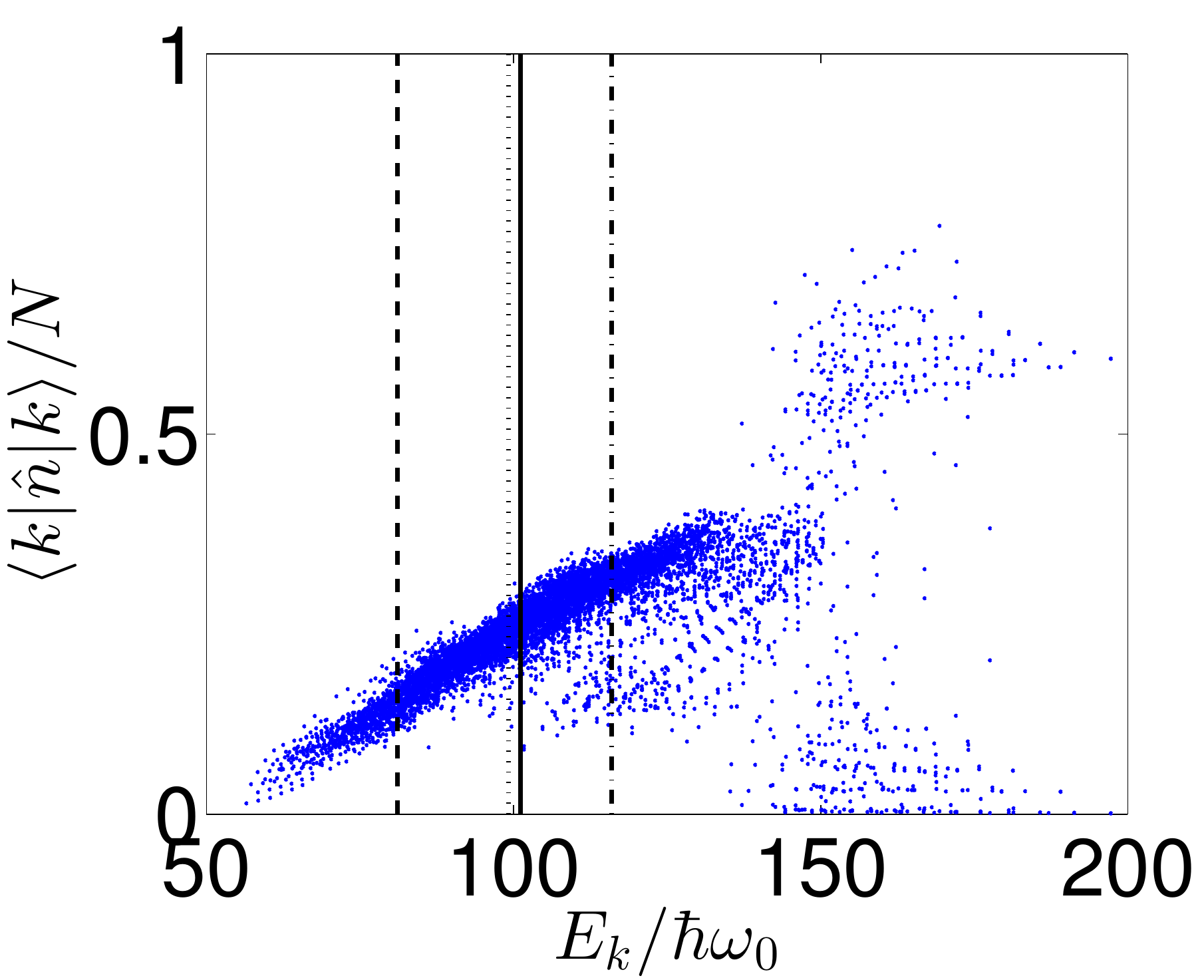}}
\protect\caption{(Color online) Distributions of EEV of $\hat{n}^0_L$ (left) and $\hat{n}^1_L$ (right) for [(a),(b)] $NU^{0}/\hbar\omega_0=2$, [(c),(d)] $NU^{0}/\hbar\omega_0=3$, and [panels (e),(f)] $NU^{0}/\hbar\omega_0=4$. Vertical lines mark the mean energies of initial states for the dynamics in 
the subsequent sections: (dashed) $E_0/\hbar\omega_0=81.053$,  (dotted) $E_0/\hbar\omega_0=99.202$, (solid) $E_0/\hbar\omega_0=101.181$, and (dashed-dotted) $E_0/\hbar\omega_0=115.9409$.
}
\label{eev2} 
\end{figure}

We plot the distribution of EEV for different interaction strengths $NU^0/\hbar\omega_0$ and finite integrability breaking $U^{01}=U^0/2$ in Fig. \ref{eev2}. 
For $NU^0/\hbar\omega_0=2$, the distribution of EEV is still regular and structured similar to the integrable case. Even though the EEV starts to clump together, separated bands are still noticeable, especially in low energies. The behavior of the distribution of EEV, in particular its smoothness, improves as the interaction strength is increased from $NU^0/\hbar\omega_0=2$ to $NU^0/\hbar\omega_0=3$. 
For $NU^0/\hbar\omega_0=4$, the vertical width of the distribution of EEV in the lower half of the spectrum is narrower than that in the remaining half. 
Note, however, that eigenstates that are very close to the ground-state energy will always violate the ETH, at least for interactions considered here, since the distribution is quite sparse in this region. This implies the absence of thermalization for initial energies close to the ground state. We point out that relaxation close to an infinite temperature state can be considered as a borderline case for which the ETH is still satisfied, as seen from the solid vertical line in Figs. \ref{eev2}(e) and \ref{eev2}(f). 
\begin{figure}[!ht]
  \includegraphics[width=0.7\columnwidth]{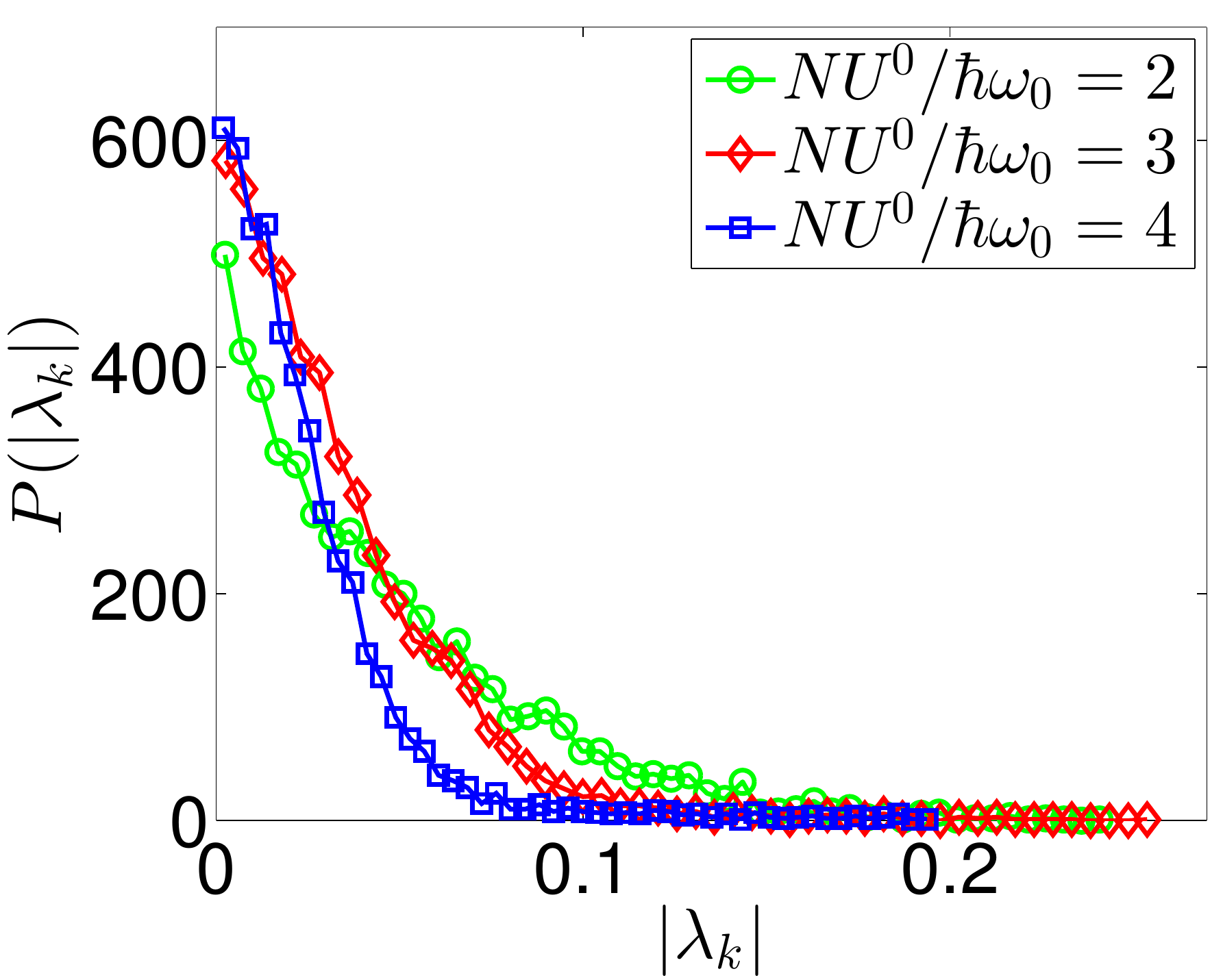}
\protect\caption{(Color online) Distributions of consecutive EEV gaps of $\hat{n}^0_L$ for different interaction strengths.
}
\label{eigdist} 
\end{figure}

A more quantitative description of the smoothness of the EEV can be drawn from the distribution of gaps between consecutive EEV \cite{Kim2014},
\begin{equation}
 |\lambda_k| = \biggl| \langle k+1|\hat{O}|k+1 \rangle - \langle k|\hat{O}|k \rangle \biggr|.
\end{equation}
If EEV is a smooth function of the eigenenergies, the distribution of $|\lambda_k|$ denoted by $P(|\lambda_k|)$ will be sharply peaked near $|\lambda_k|=0$. Accordingly, the width of $P(|\lambda_k|)$ decreases as the distribution of EEV smoothens. It is clear from Fig.~\ref{eev2} that the ETH works only for the bulk of the spectrum, i.e., in the lower half of the spectrum. Similar to Ref.~\cite{Kim2014}, we consider only a part of the full spectrum by removing the lowest $10 \% $ and the highest $25 \%$ of the eigenstates. In Fig.~\ref{eigdist}, we plot $P(|\lambda_k|)$ for the rescaled occupation number in the lower mode of the left well $\hat{n}^0_L/\sqrt{N}$. As we decrease the interaction strength, the distribution of $|\lambda_k|$ broadens and the peak decreases. This is consistent with the qualitative picture presented in Fig.~\ref{eev2}.

Our results in this section suggest that despite the final Hamiltonian being near integrability, the fact that it is still nonintegrable permits the fulfillment of the ETH but not for all eigenstates. Therefore, the condition for an initial state to properly thermalize will largely depend on which part of the spectrum the energy of an initial state is located. This is indeed the case, as we see later.

\section{Quench Dynamics}\label{sec:quenchdyn}
\subsection{Comparison of diagonal to microcanonical expectation values}\label{sec:deme}

We are interested in the relaxation dynamics if the system starts out as a product state of each mode. 
At long times, local operators in the system are expected to fluctuate around the mean value predicted by the diagonal ensemble
\begin{equation}\label{eq:dens}
	\overline{\langle \hat{A} \rangle} = \sum_k |C^k_{n_0}|^2 A_{kk},
\end{equation}
where $A_{kk}=\langle k | \hat{A} |k \rangle$.
Clearly, the diagonal ensemble will depend on the initial state through $|C^k_{n_0}|^2$. On the other hand, the microcanonical average depends only on the initial energy and for finite system will also be a function of the energy window $\Delta E$. The energy window is chosen such that it is small enough to approximate the mean energy of the system but still large enough to contain significant number of eigenstates. 
The appropriate microcanonical prediction of a local operator is obtained by averaging over all eigenstates within a small window of energy $[E_0-\Delta E,E_0+\Delta E]$
\begin{equation}\label{eq:mens}
	\langle A \rangle_{\mathrm{ME}}=\frac{1}{\mathcal{N}_{E,\Delta E}}\sum_{|E_k-E_0|<\Delta E}A_{kk},
\end{equation}
where $E_0=\langle n_0|\hat{H}|n_0 \rangle$ is the initial energy of the system and ${\mathcal{N}_{E,\Delta E}}$ corresponds to the number of eigenstates within the chosen window.

\begin{figure}[!ht]
\includegraphics[width=0.7\columnwidth]{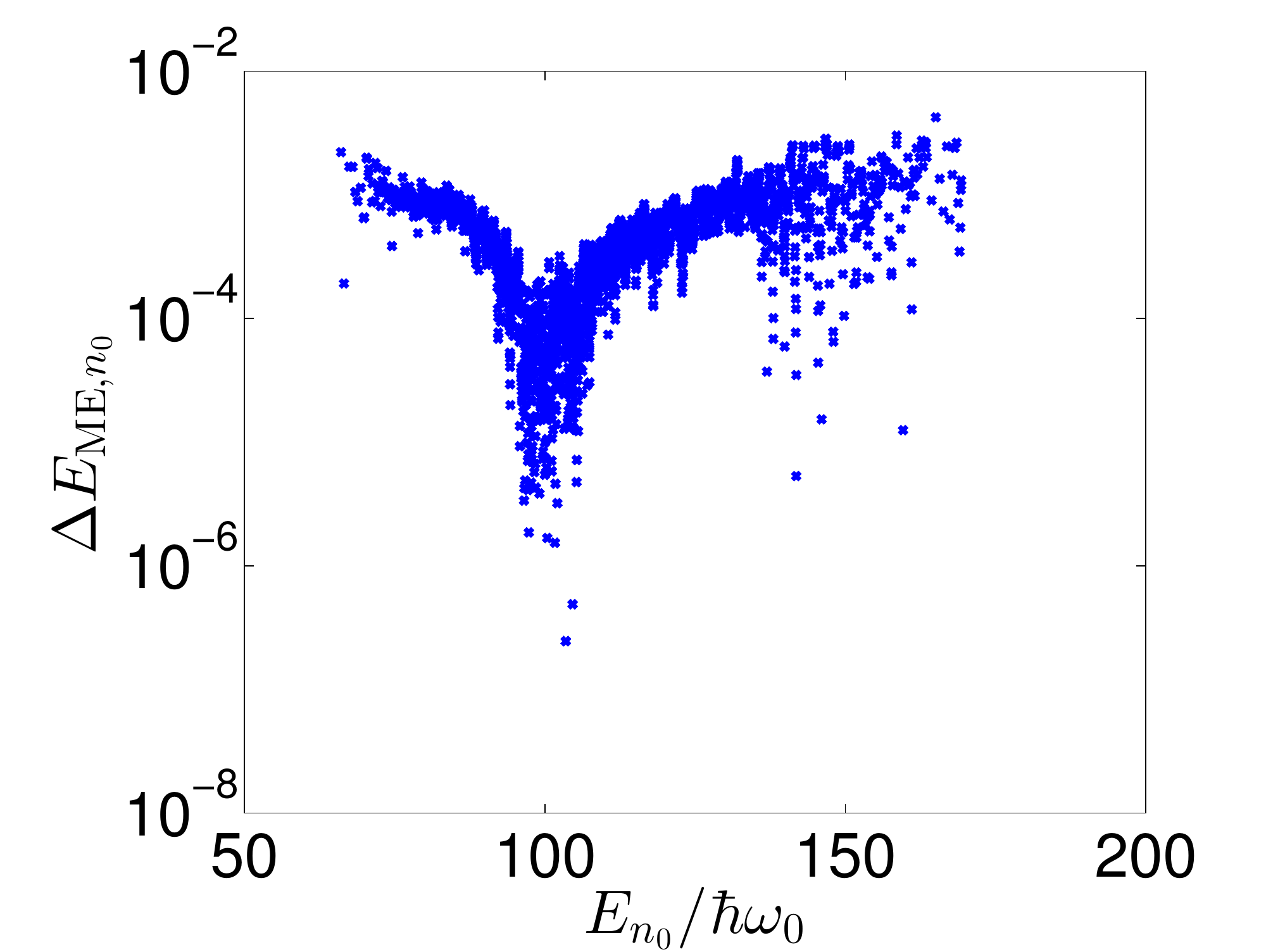}
\protect\caption{(Color online) Relative deviations between the microcanonical and the mean energy of all possible initial Fock states for $NU^0/\hbar\omega_0=4$.}
\label{fig:mc} 
\end{figure}

Normally, for every initial state, one has to first calculate $\langle\hat{H}\rangle_{\mathrm{ME}}=E_{\mathrm{ME}}$ over a range of $\Delta E$ and then choose $\Delta E$ that best approximates $E_0$.
Here, in order to ease up on computational expense, we simply fix the energy window by choosing $\Delta E= 0.02 E_0$. Although we checked that the microcanonical averages obtained this way are still weakly dependent on other choices of $\Delta E$, to keep track of how close the microcanonical energy is to the energy of the system, we define the relative deviation 
\begin{equation}
	\Delta E_{\mathrm{ME},n_0} = \frac{|E_0-E_{\mathrm{ME}}|}{E_0}.
\end{equation}
The plot presented in Fig. \ref{fig:mc}, which shows the relative deviations of microcanonical averages in the case $NU^0/\hbar\omega_0=4$, underpins our choice of $\Delta E$.

\begin{figure}[!ht]
\subfloat[]{\includegraphics[width=0.5\columnwidth]{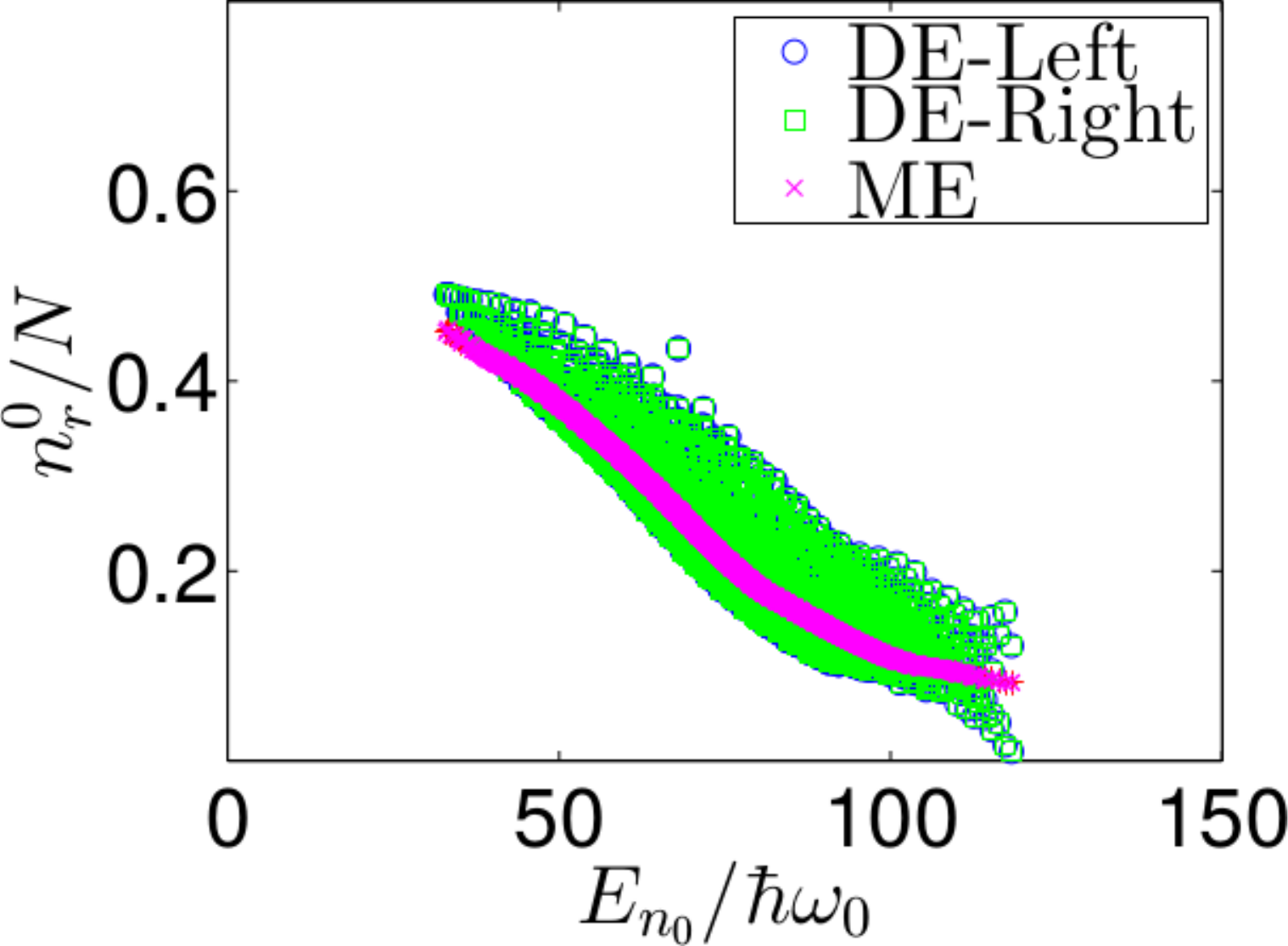}}\subfloat[]{\includegraphics[width=0.5\columnwidth]{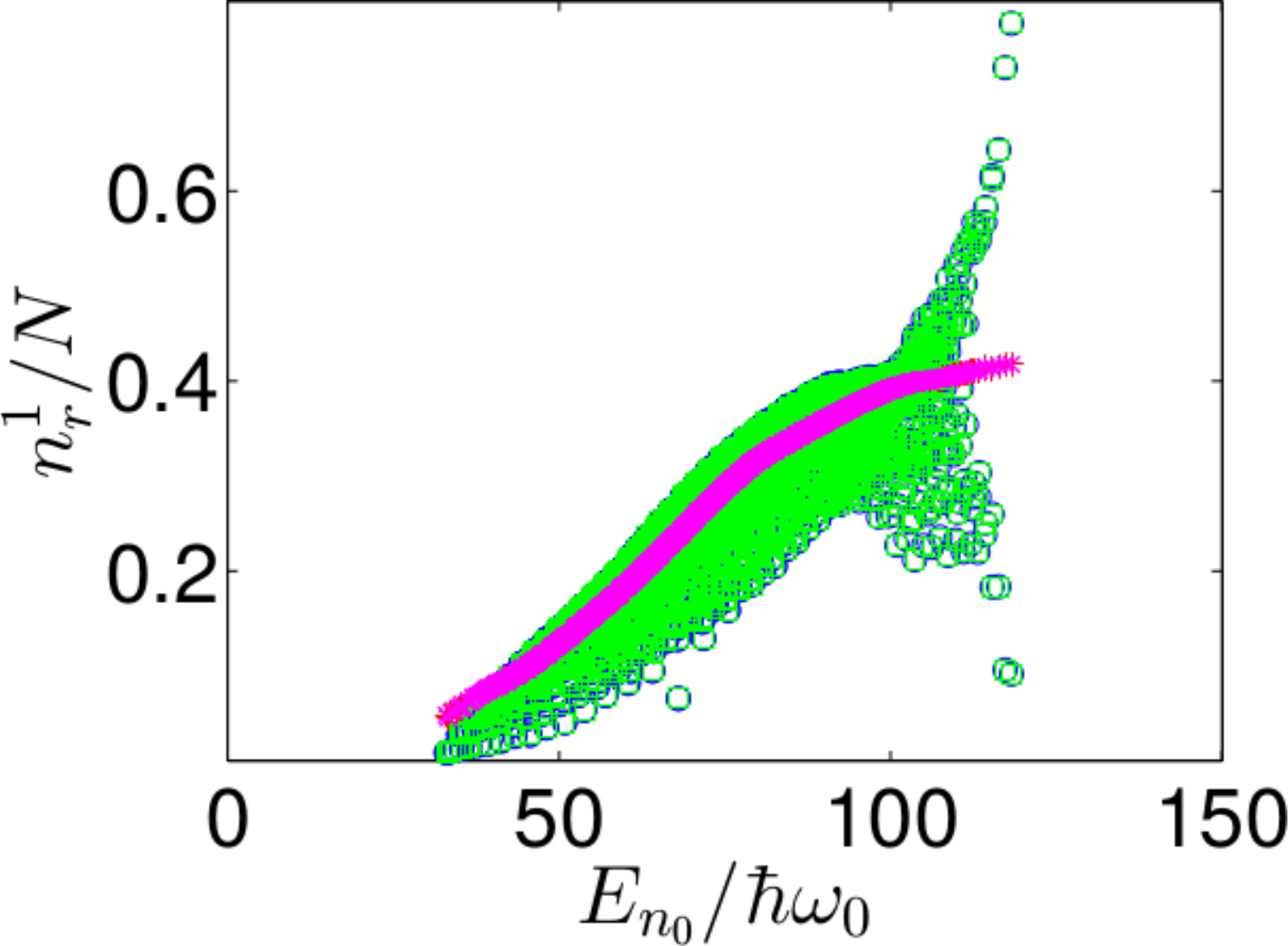}}\\
\subfloat[]{\includegraphics[width=0.5\columnwidth]{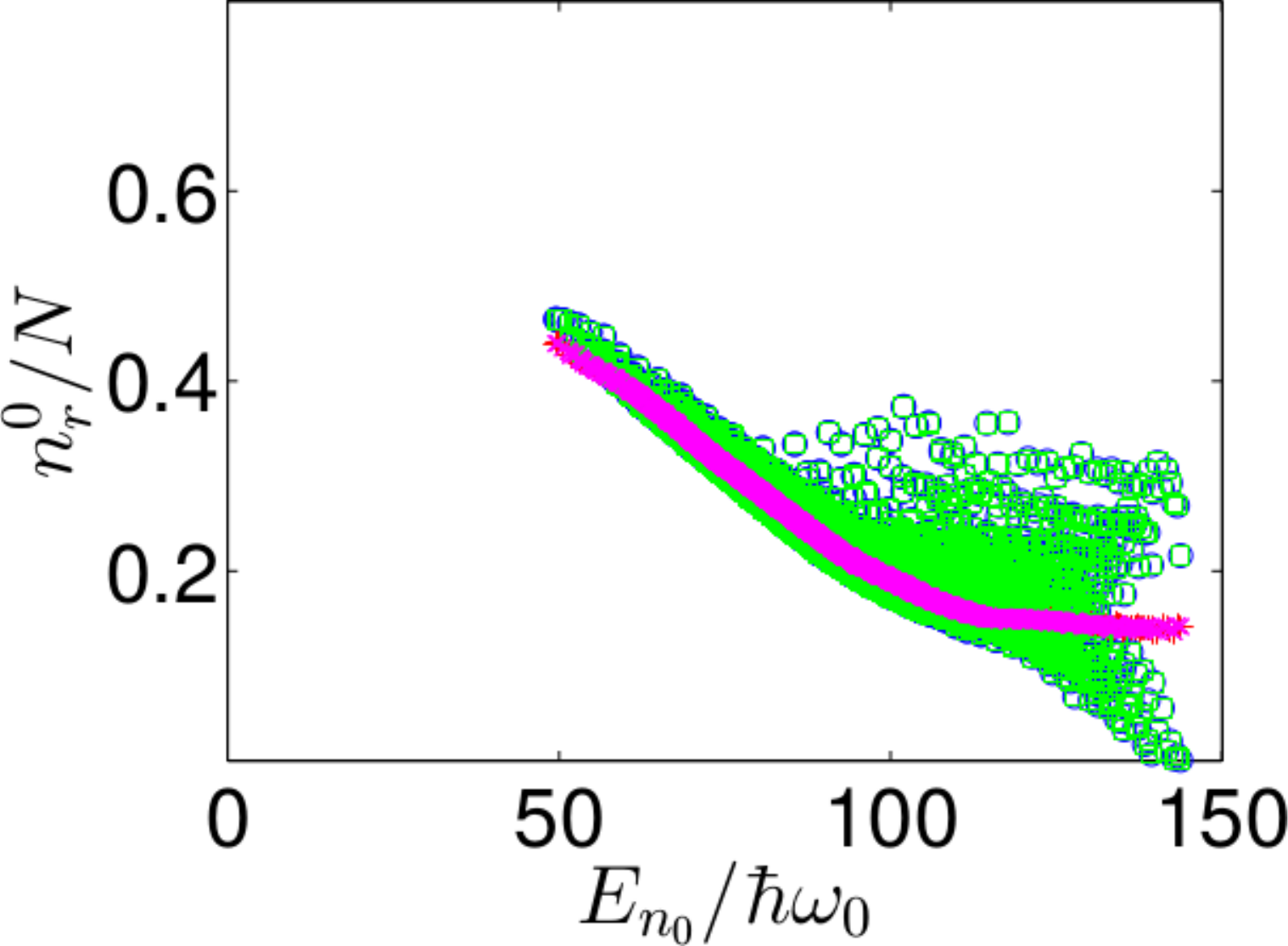}}\subfloat[]{\includegraphics[width=0.5\columnwidth]{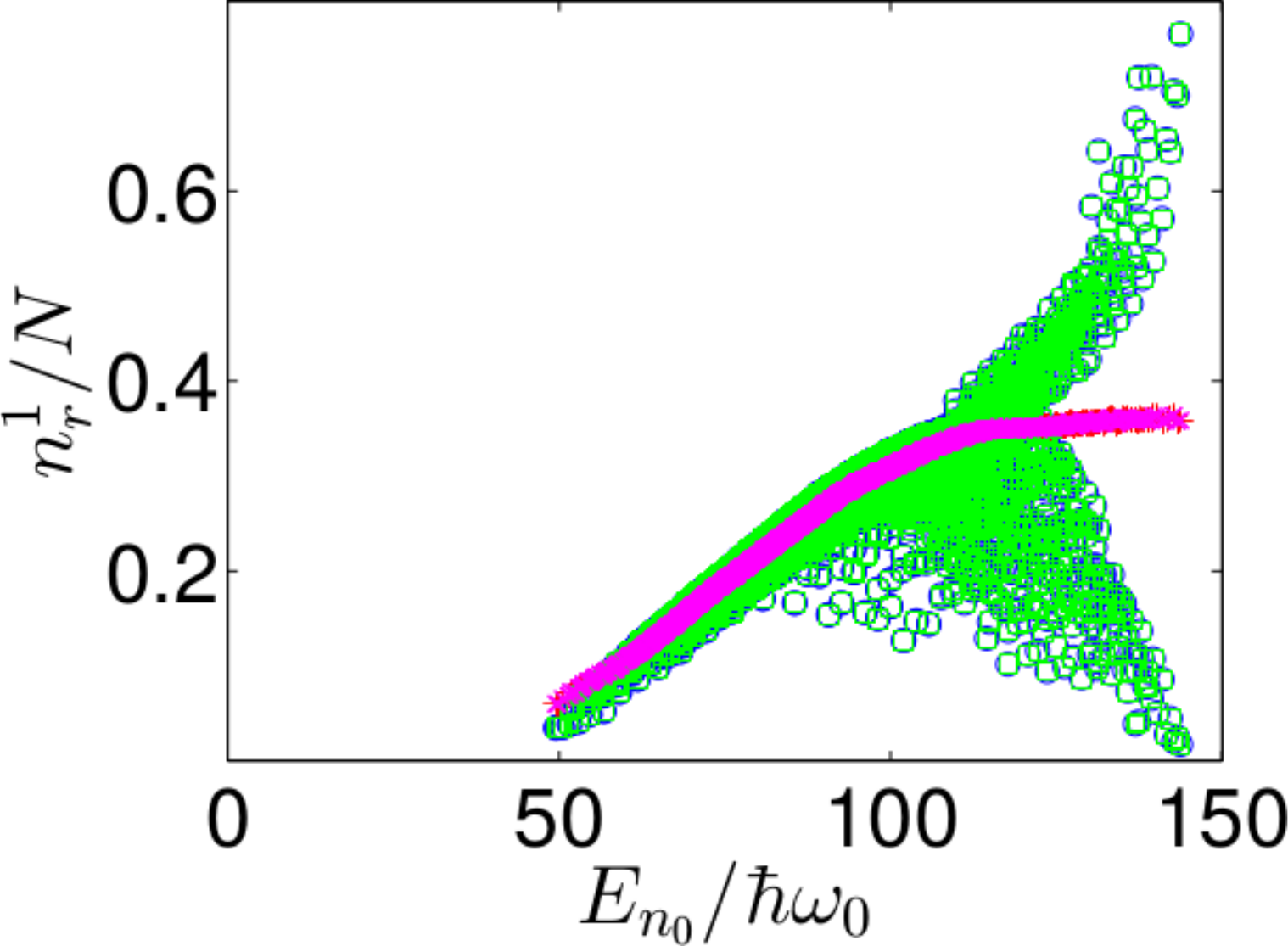}}
\protect\caption{(Color online) Diagonal ensemble (DE) vs microcanonical ensemble (ME) averages of local operators for all possible initial Fock states. (a) $\hat{n}^0_r$ for $NU^{0}/\hbar\omega_0=2$; (b) $\hat{n}^1_r$ for $NU^{0}/\hbar\omega_0=2$; (c) $\hat{n}^0_r$ for $NU^{0}/\hbar\omega_0=3$;  and (d) $\hat{n}^1_r$ for $NU^{0}/\hbar\omega_0=3$.}
\label{fig:diag1} 
\end{figure}

\begin{figure}[!ht]
\subfloat[]{\includegraphics[width=0.5\columnwidth]{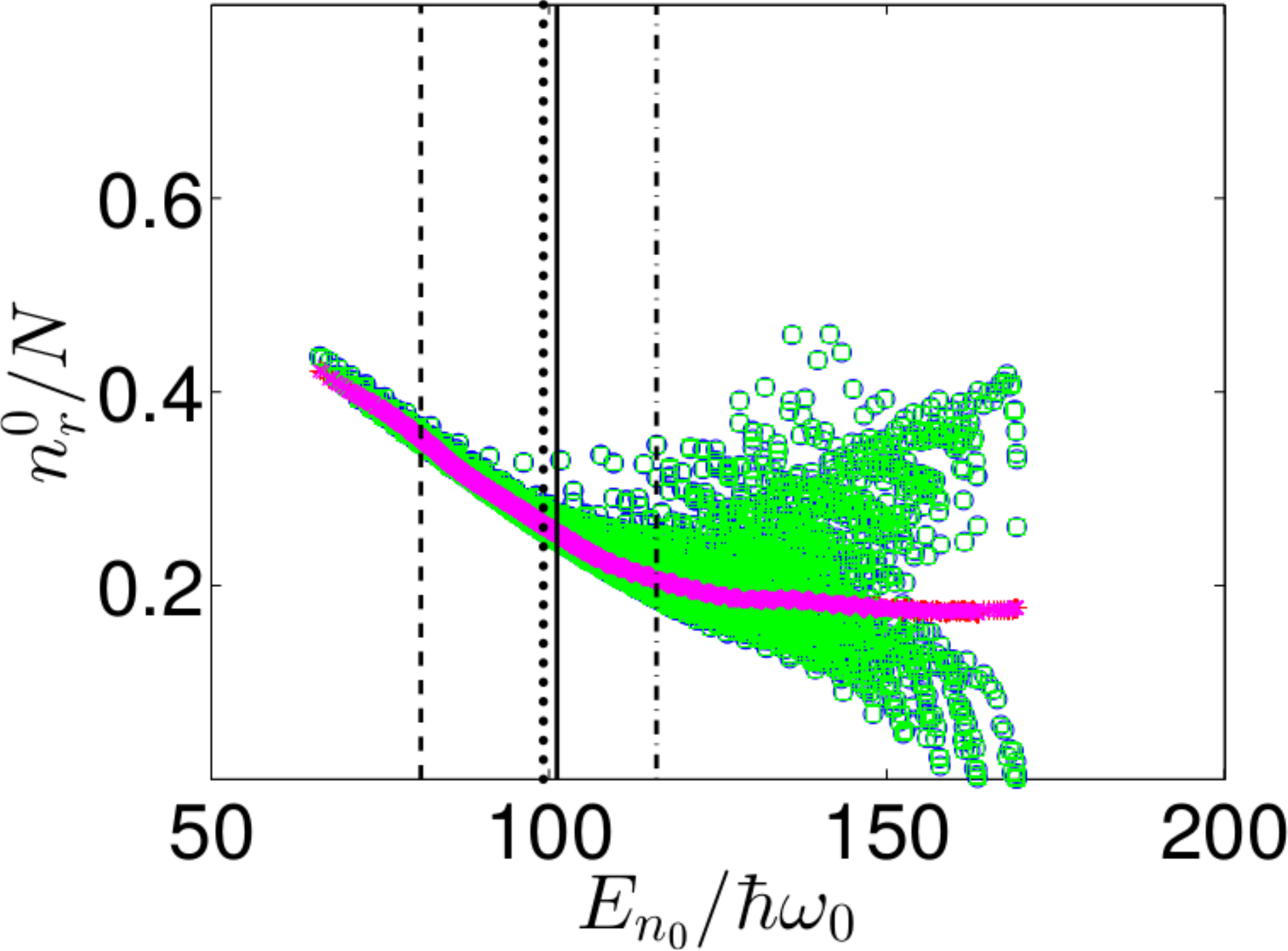}}\subfloat[]{\includegraphics[width=0.5\columnwidth]{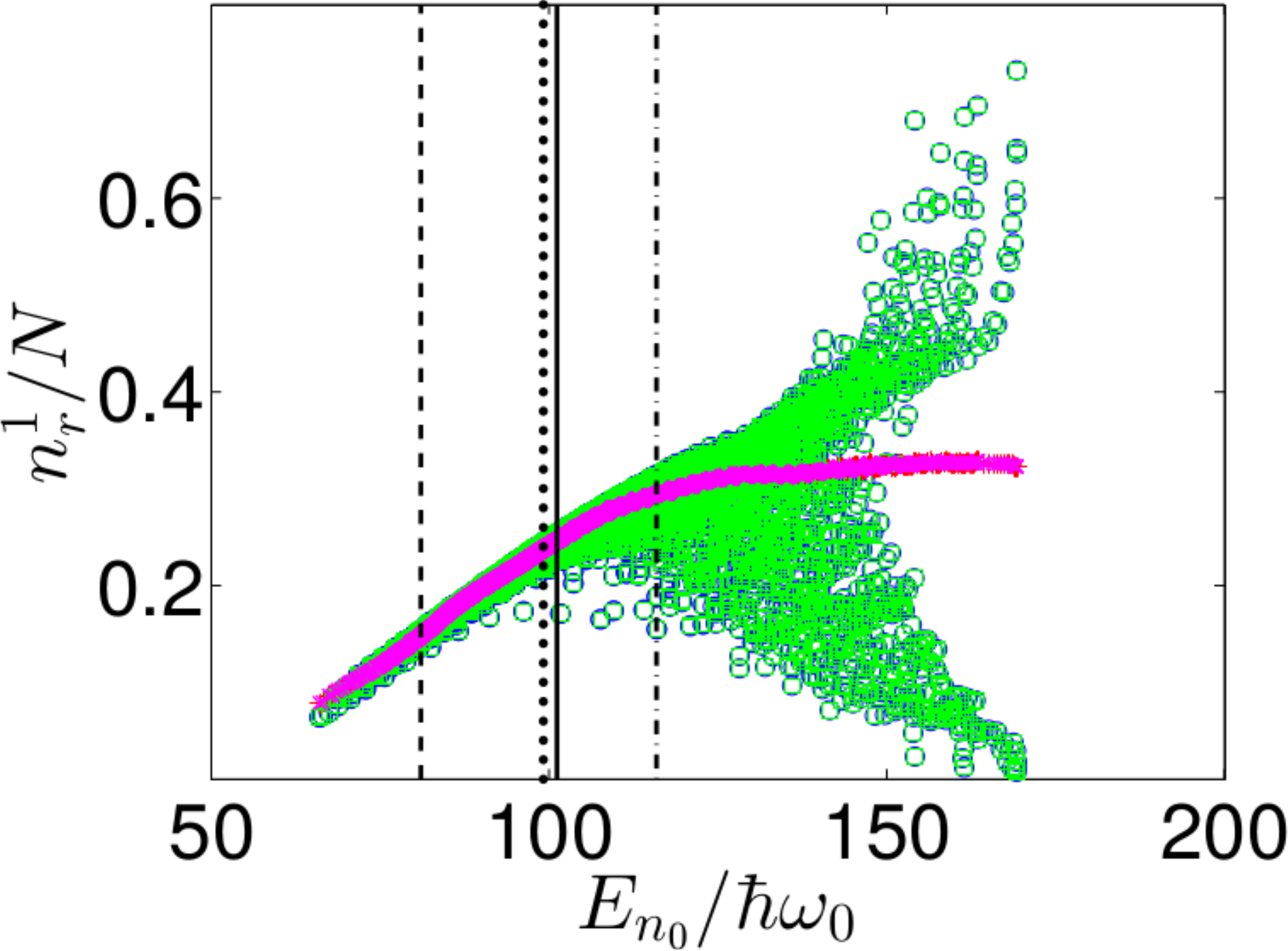}}\\
\subfloat[]{\includegraphics[width=0.5\columnwidth]{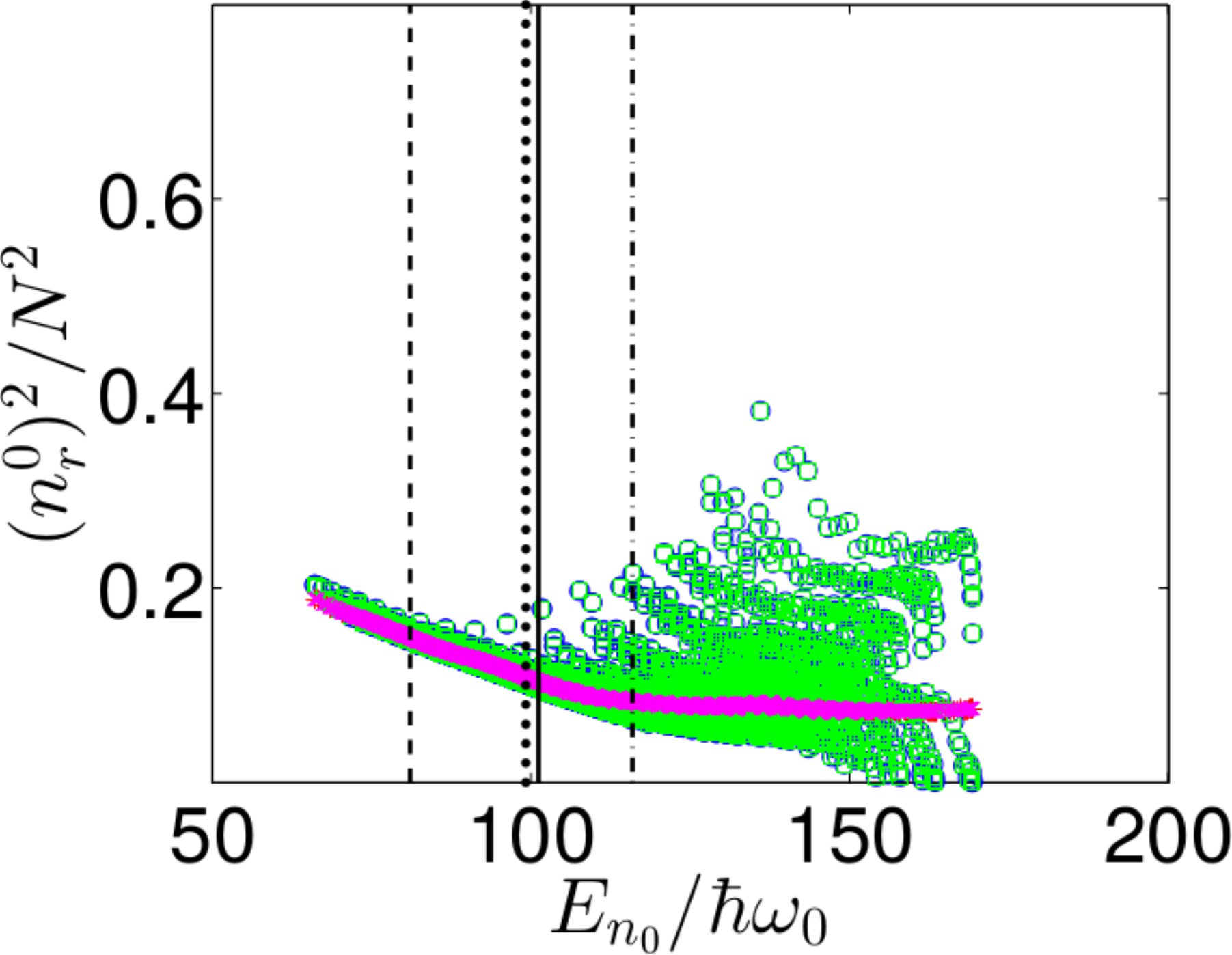}}\subfloat[]{\includegraphics[width=0.5\columnwidth]{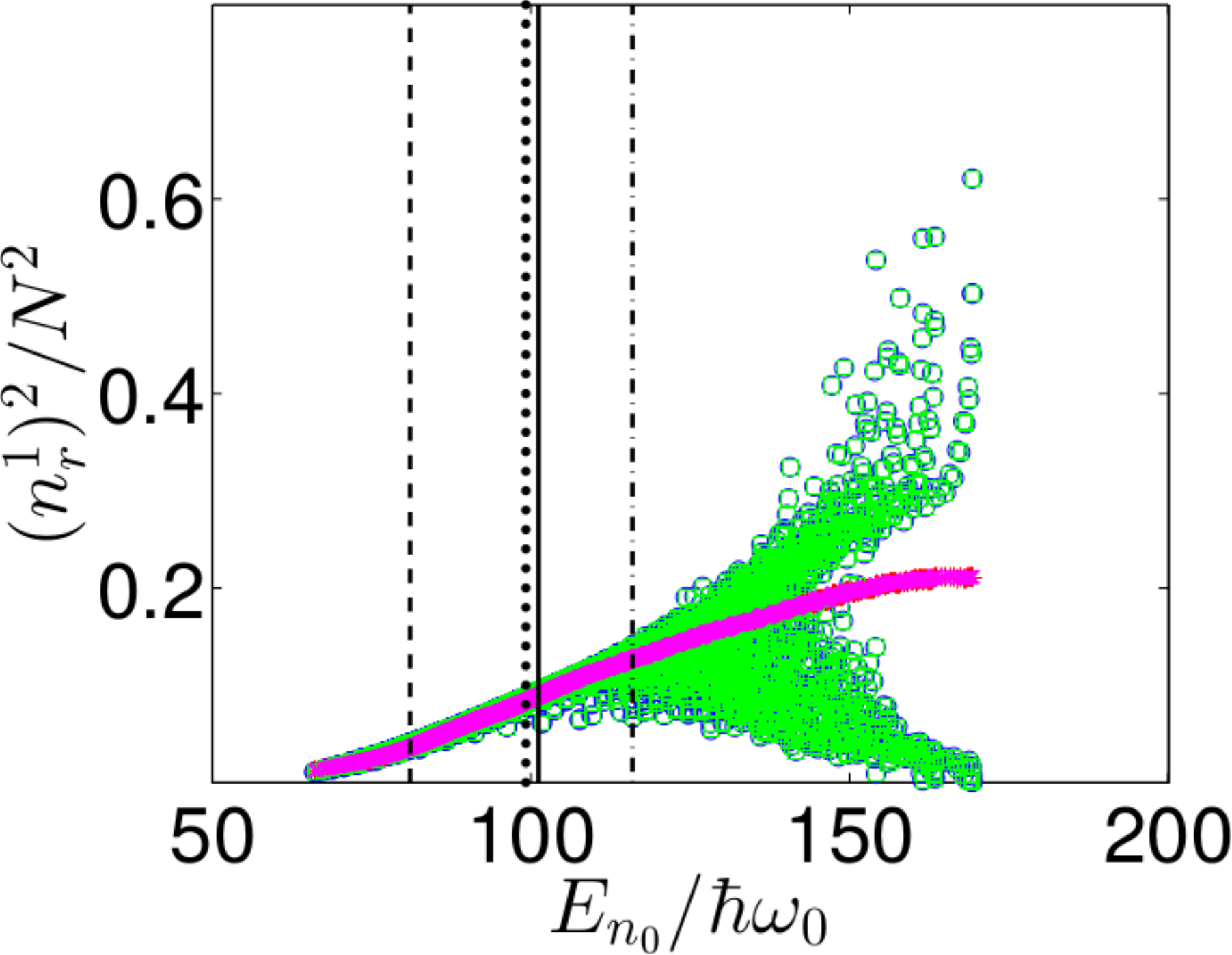}}
\protect\caption{(Color online) Similar to Fig. \ref{fig:diag1} but $NU^{0}/\hbar\omega_0=4$. (a) $\langle \hat{n}^0_r \rangle$; (b) $\langle \hat{n}^1_r \rangle$; (c) $\langle (\hat{n}^0_r)^2 \rangle$; (d) $\langle (\hat{n}^1_r)^2 \rangle$. Vertical lines mark the energies similar to Fig.~\ref{eev2}.}
\label{fig:diag2} 
\end{figure}

The postquench Hamiltonian Eq. \eqref{eq:Hamilt} has parity or reflection symmetry about the center of the trap and so a thermalized state is expected to respect this symmetry. That is, the stationary properties in the left well must be the same as those in the right well. This is indeed the case for both the microcanonical and the diagonal ensemble predictions shown in Figs. \ref{fig:diag1} and \ref{fig:diag2}. In each plot, both ensemble averages in the left well are sitting on top of the corresponding values in the right well.

In Figs. \ref{fig:diag1} and \ref{fig:diag2}, we compare the microcanonical and the diagonal ensemble values calculated using Eqs.~\eqref{eq:dens} and \eqref{eq:mens}, respectively, for all initial Fock state configurations. Similar to the distribution of the EEV, thermalization is only viable if the distribution of the diagonal ensemble averages is smooth and monotonous across the range of possible initial energies. This is an alternative way of describing another signature of thermalization, which is the independence of the equilibrium state on the details of initial states. The results shown in Figs.~\ref{fig:diag1} and \ref{fig:diag2} are consistent with the prediction of thermalization in the context of the ETH as discussed in Sec. \ref{sec:eth}. 
That is, the microcanonical ensemble works well in predicting the long-time average of an initial state with energy somewhere in the smooth region of the EEV distribution.
Recall that in Figs.~\ref{eev2}(a) and \ref{eev2}(b), the vertical width of the EEV is large for $NU^0/\hbar\omega_0=2$. This explains why the microcanonical averages are drastically different from the diagonal ensemble values for most of the initial states shown in Figs.~\ref{fig:diag1}(a) and \ref{fig:diag1}(b). Following the same line of reasoning, we find improvements in the behavior of the diagonal ensemble and its agreement with the microcanonical average when the interaction strength is increased to $NU^0/\hbar\omega_0=3$ and $NU^0/\hbar\omega_0=4$. These are shown in Figs.~\ref{fig:diag1}(c), \ref{fig:diag1}(d), \ref{fig:diag2}(a), and \ref{fig:diag2}(b).
We stress that there is a discrepancy found between the microcanonical and the long-time predictions for initial states with sufficiently low energies, in agreement with observations drawn from the EEV near the ground state. 

Aside from the local one-body operator $\hat{n}^{\ell}_r$, we also calculate the ensemble averages for the local two-body operator $\langle (\hat{n}^{\ell}_r)^2 \rangle$. We plot the results in  Figs.~\ref{fig:diag2}(c) and \ref{fig:diag2}(d). This quantity is essential in calculating the average energy of a mode, which will be further investigated in Sec. \ref{sec:sub}.

\subsection{Delocalized initial states}\label{sec:chaos}

In the spirit of the energy shell approach, consider a set of unperturbed basis states $|n\rangle$. When a finite interaction is turned on, this perturbation will couple some of these states $|n\rangle$. The exact eigenstates of the final Hamiltonian can be expressed as superpositions of the unperturbed states $|k \rangle = \sum_n C^{k}_n |n\rangle$. It was conjectured that the overlap coefficients $C^{k}_n$ become random variables from a Gaussian distribution defined by the ``energy shell'' \cite{Casati1993, Casati1996}.
For systems with only two-body interaction and sufficiently strong perturbations, regardless of integrability, the shape of the energy shell follows a Gaussian profile centered at the initial mean energy and its width is equal to the energy variance \cite{Santos2012}.
Statistical description of the system is viable if the shape of the local density of states (LDoS) or strength function resembles a Gaussian of the same mean and variance as the energy shell \cite{Santos2012c}.
Recently, it was conjectured that for initial states in the middle of the spectrum, thermalization is guaranteed if the initial state ergodically fills the area defined by the energy shell, irrespective of integrability (or chaoticity) of the final Hamiltonian \cite{Torres2013}. These initial states are referred to as chaotic, which can be viewed as delocalization of the initial unperturbed state with respect to the eigenstates of the final Hamiltonian, although, whether this phenomenon can occur for other models and more importantly for initial states not too close to the center of the spectrum \cite{Torres2013} is still an open question. Here, we respond to this query by demonstrating delocalization of initial states with energy away from the middle of the spectrum even though our postquench Hamiltonian is far from having a chaotic spectrum.

The LDoS or energy distribution is defined by the distribution of $|C^k_{n_0}|^2$ in the eigenvalues $E_k$
\begin{equation}\label{LDoS}
 \mathcal{F}_{n_0}(E) = \sum_k |C^k_{n_0}|^2\delta(E-E_k).
\end{equation}
In practice, the LDoS is numerically obtained by dividing the whole spectrum of the final Hamiltonian and
then taking $\sum_k |C^k_{n_0}|^2$ in each bin \cite{Torres2014b}. 
In the limit of strong perturbation, the shape of the LDoS for finite closed system with two-body interactions follows a Gaussian
distribution defined by the energy shell centered at $E_0$ and width $\sigma = \sqrt{\sum_k |C^k_{n_0}|^2(E_k-E_0)^2}$ (see Refs. \cite{Flambaum2001, Izrailev2006, Torres2014}, and references therein),
\begin{equation}
  \mathcal{F}_G(E) = \frac{1}{\sqrt{2\pi\sigma^2}}\mathrm{exp}\biggl[-\frac{(E-E_0)^2}{2\sigma^2}\biggr].
\end{equation}

We now investigate delocalization of initial states in the eigenstates of the final Hamiltonian and check if it coincides with the onset of thermalization found in the previous sections. We focus on the case $NU^0/\hbar\omega_0=4$.
\begin{figure}[!htb]
\subfloat[]{\includegraphics[width=0.51\columnwidth]{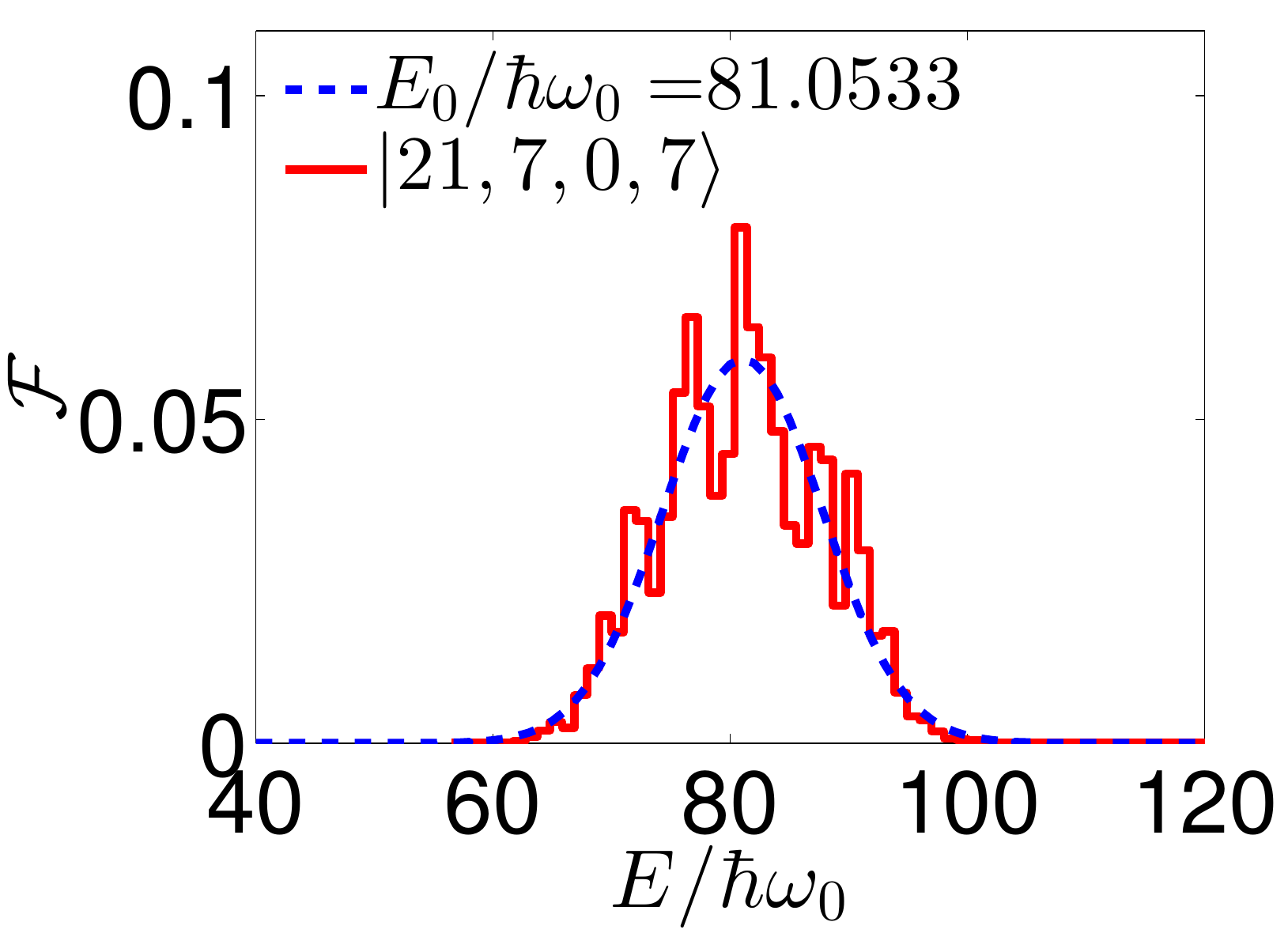}}\subfloat[]{\includegraphics[width=0.49\columnwidth]{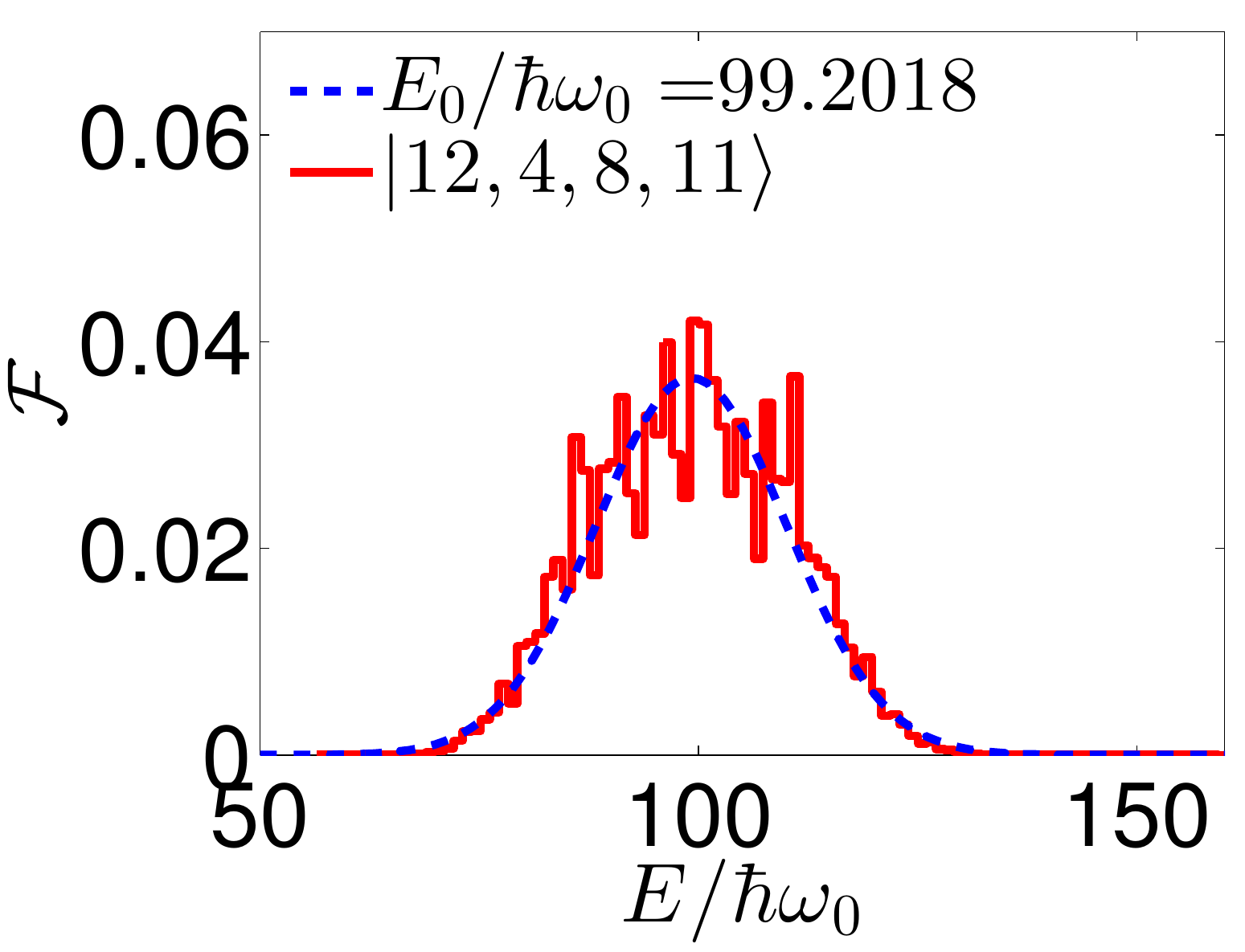}}\\
\subfloat[]{\includegraphics[width=0.49\columnwidth]{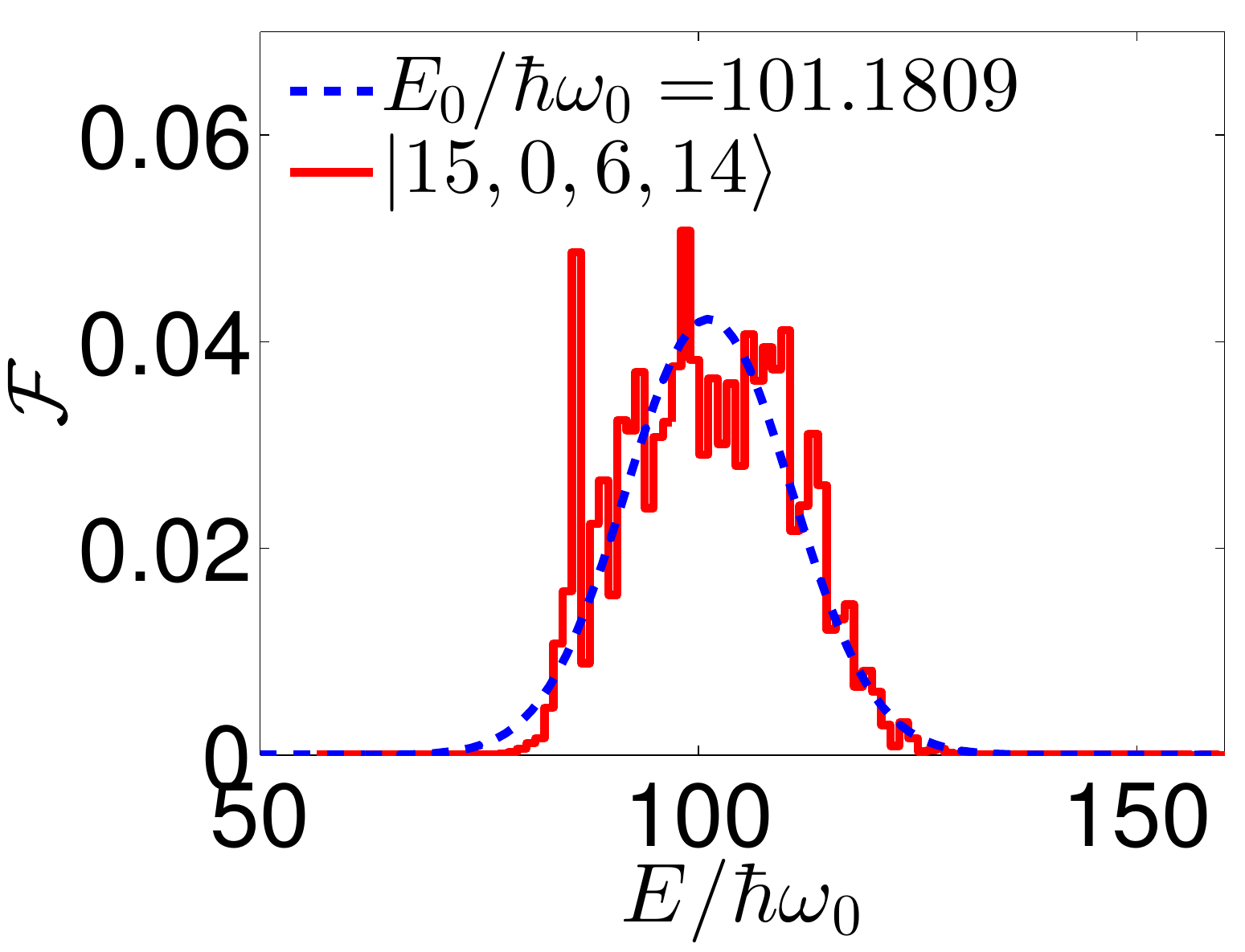}}\subfloat[]{\includegraphics[width=0.51\columnwidth]{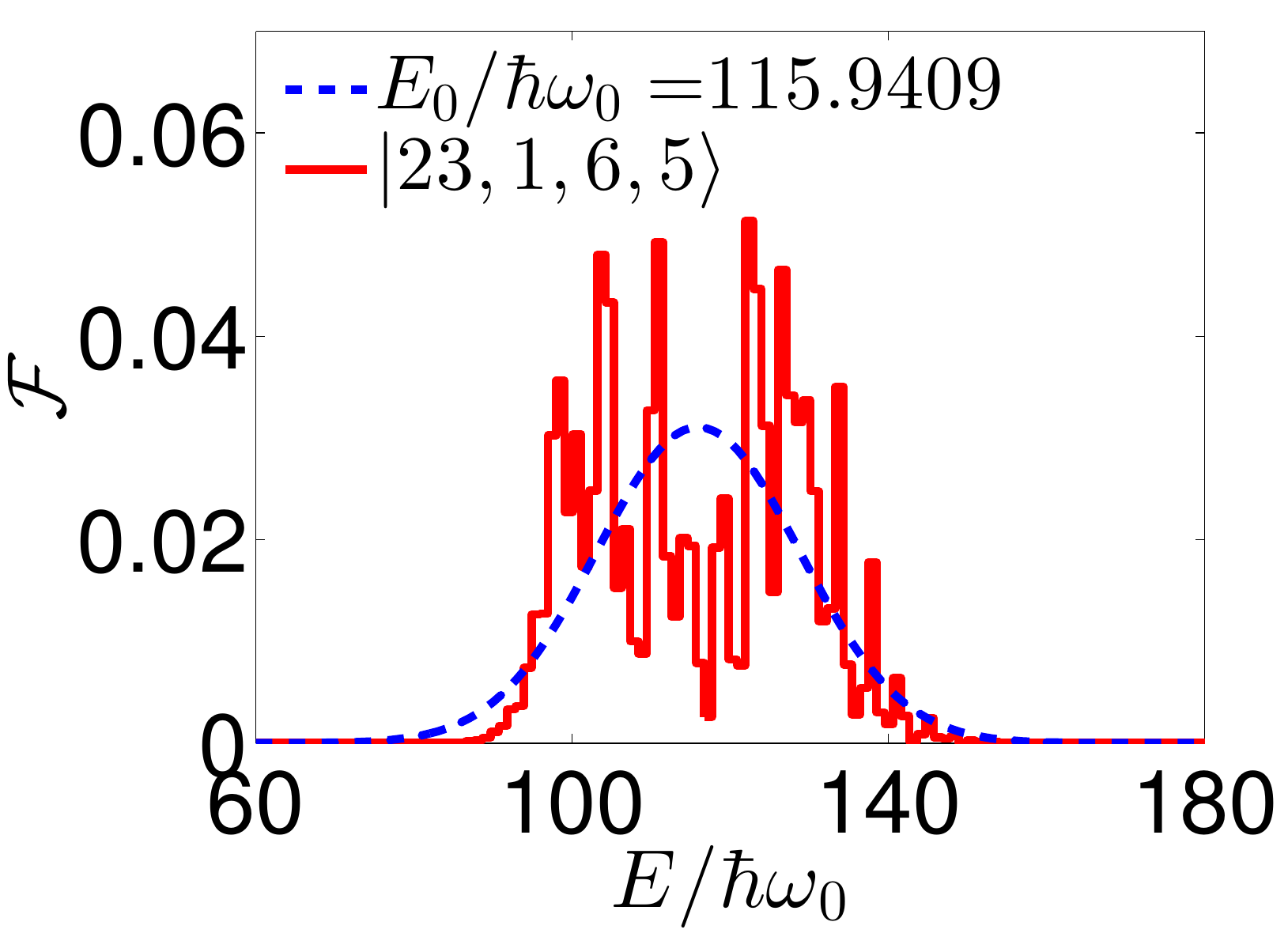}}
\protect\caption{(Color online) LDoS of different initial states for $NU^{0}/\hbar\omega_0=4$. The dashed lines correspond to the energy shell.}
\label{LDoSdec}
\end{figure}
In Figs.~\ref{eev2} and \ref{fig:diag2}, we find smooth and monotonic behavior of the EEV and diagonal ensemble average in the lower half of the spectrum
while the remaining half suffers from strong fluctuations. Following this observation, we choose four different initial energies $E_0/\hbar\omega_0 = \{81.05, 99.20, 101.18,115.94\}$.
Specifically, we choose representative initial Fock states with the aforementioned set of energies indicated by the vertical lines in Figs.~\ref{eev2} and \ref{fig:diag2}.

We use 140 bins in the numerical calculation of LDoS shown in Fig. \ref{LDoSdec}.
We determine whether the initial state is chaotic with respect to the exact eigenstates by checking if the shape of its LDoS is close to the shape of the energy shell. 
In Figs.~\ref{LDoSdec}(a)-\ref{LDoSdec}(c), we confirm the existence of chaotic initial states in the system by showing that for these initial states, most areas defined by the energy shell are nicely filled.
More importantly, these delocalized initial states display thermal behavior as shown in Fig. \ref{fig:diag2}. Due to the system being finite, strong fluctuations around the mean energy is visible. Nevertheless, the behavior in the tails of the energy distribution is still quite close to the Gaussian tail of the energy shell. This is consistent with observations in other
finite systems such as those found in Ref. \cite{Santos2011}. 

Also shown in Fig. \ref{LDoSdec}(d) is an initial state close to the middle but already in the upper half of the spectrum, $E_0/\hbar\omega_0 = 115.94$. Clearly, this state no longer has a Gaussian profile and its LDoS exhibits multiple peaks. It is possible that the energy distribution of this state can be described by a bimodal Lorentzian distribution similar to the one in Ref. \cite{Torres2014b} but a more careful numerical analysis is needed. Note that this particular case, $E_0/\hbar\omega_0 = 115.94$, reflects the failure of the ETH in the system since the distribution of the EEV for this choice of energy is quite broad, as seen in Fig. \ref{eev2}. Moreover, there are significantly stronger fluctuations in the diagonal ensemble averages of initial states around this energy as depicted in Fig. \ref{fig:diag2}. Thus, this initial state is not expected to thermalize.

\subsection{Relaxation of mode occupation number}\label{sec:relmod}

Our observable of interest that is local to a subsystem is the occupation number
of a mode $\hat{n}^{\ell}_r$.
In this section, we study the relaxation dynamics from the subsequent time evolution of the mode occupation number $\langle \hat{n}^{\ell}_r \rangle$ after the proposed quench.

\begin{figure}[!ht]
\includegraphics[width=0.7\columnwidth]{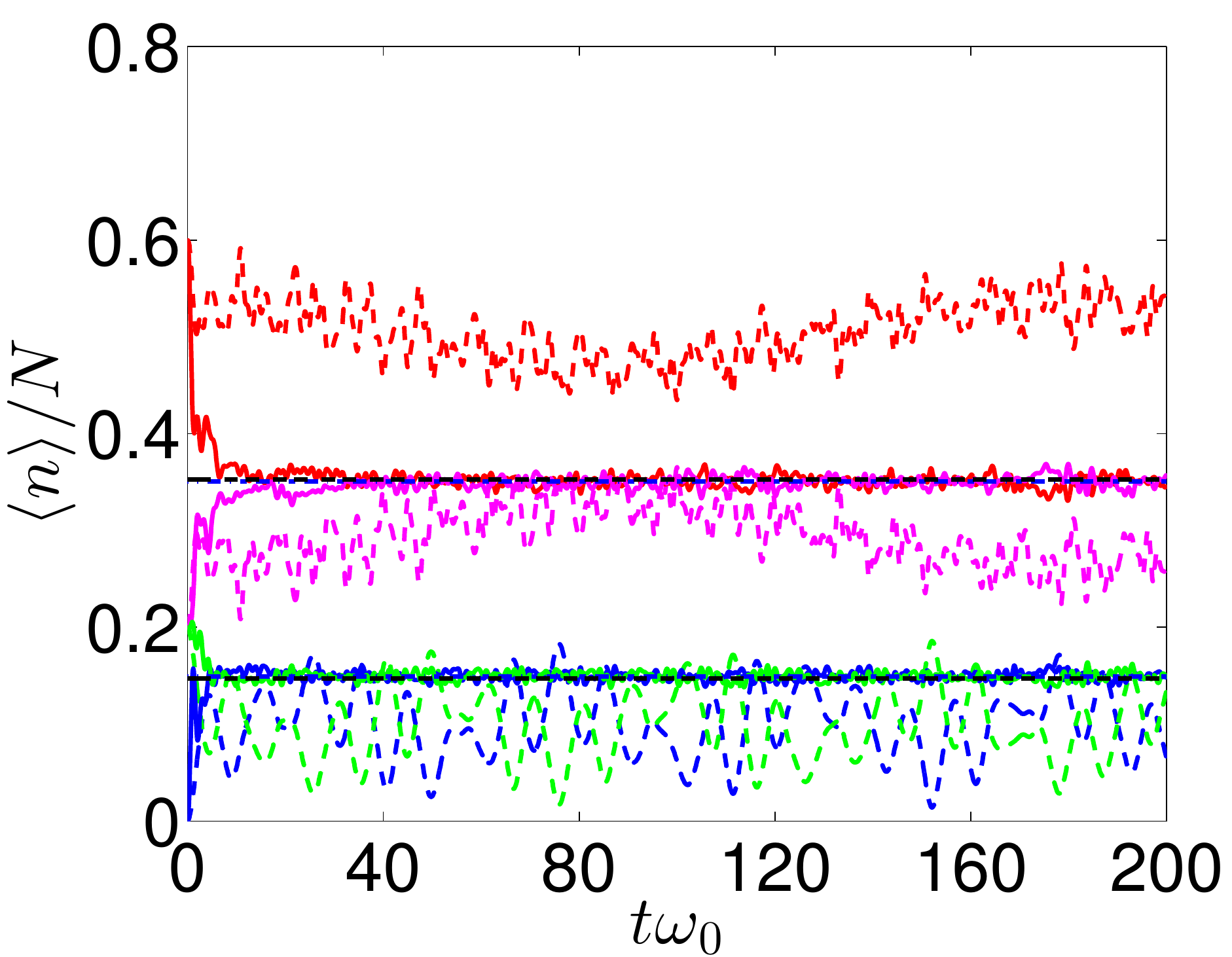}
\protect\caption{(Color online) Relaxation dynamics of occupation number for  $NU^{0}/\hbar\omega_0=4$.
Integrable case $U^{01}/\hbar\omega_0=0$ (dashed line) and finite $NU^{01}/\hbar\omega_0=2$ (solid line). The diagonal ensemble average is shown in black solid lines for $NU^{01}/\hbar\omega_0=2$. Colors denote: (red,top) $\hat{n}^0_L$, (magenta,middle) $\hat{n}^0_R$, (blue [gray], bottom) $\hat{n}^1_L$, and (green [light gray], bottom) $\hat{n}^1_R$.}
\label{fig:u0n} 
\end{figure}

In Fig. \ref{fig:u0n}, we illustrate how the dynamics of the integrable dimer $U^{01}=0$ differ from the nonintegrable case with finite interlevel coupling $NU^{01}/\hbar\omega_0=2$. The initial Fock state is the same in both cases. In addition to strong fluctuations, recurrences occur during the time evolution of the mode occupation number in the integrable case. In contrast, we observe relaxation of the mode occupation number to the value predicted by the diagonal ensemble for the nonintegrable system. Henceforth, we focus on the nonintegrable case of finite interlevel coupling $U^{01}=U^0/2$.

\begin{figure}[!ht]
\subfloat[]{\includegraphics[width=0.85\columnwidth]{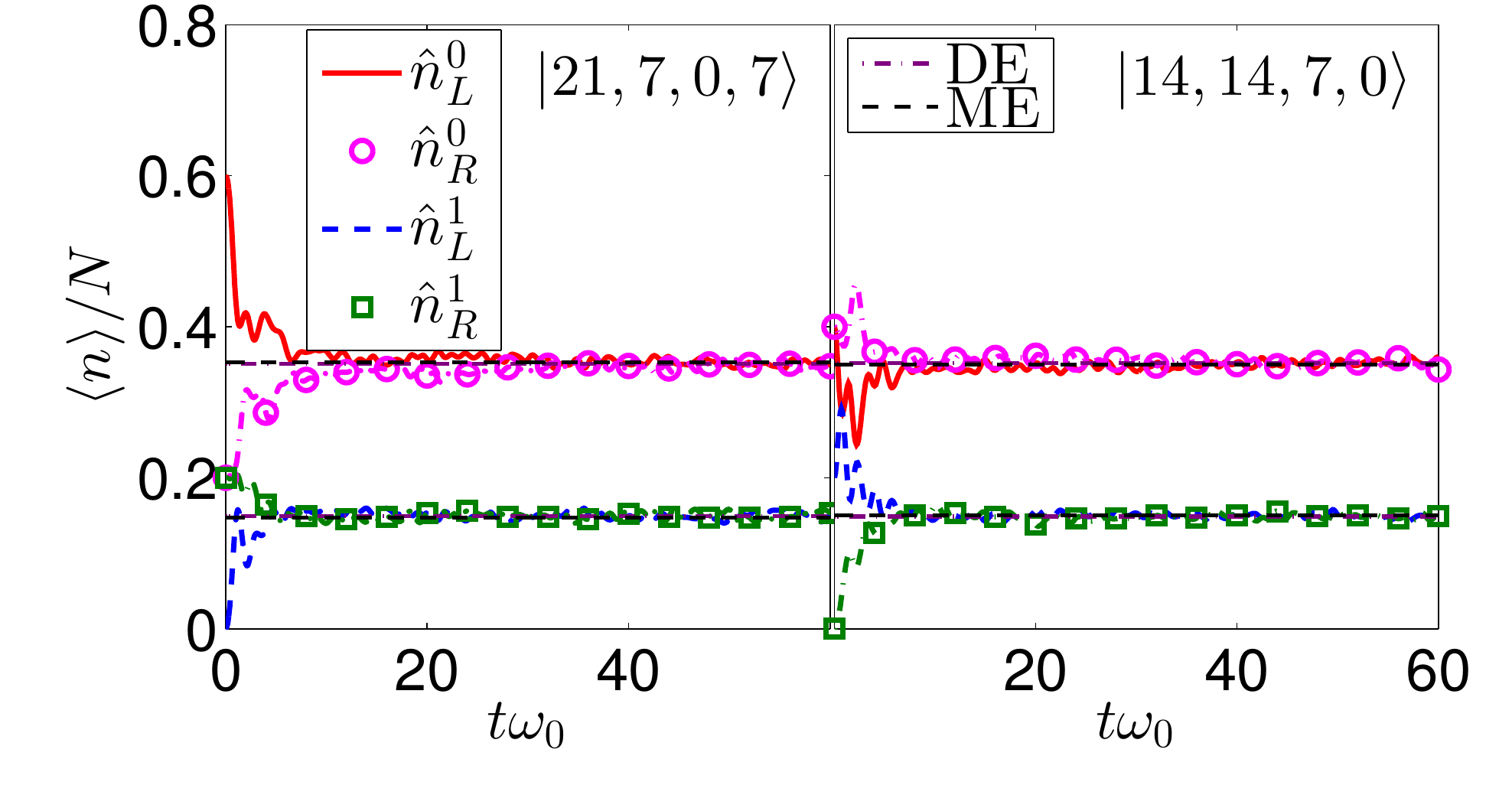}}\\
\subfloat[]{\includegraphics[width=0.85\columnwidth]{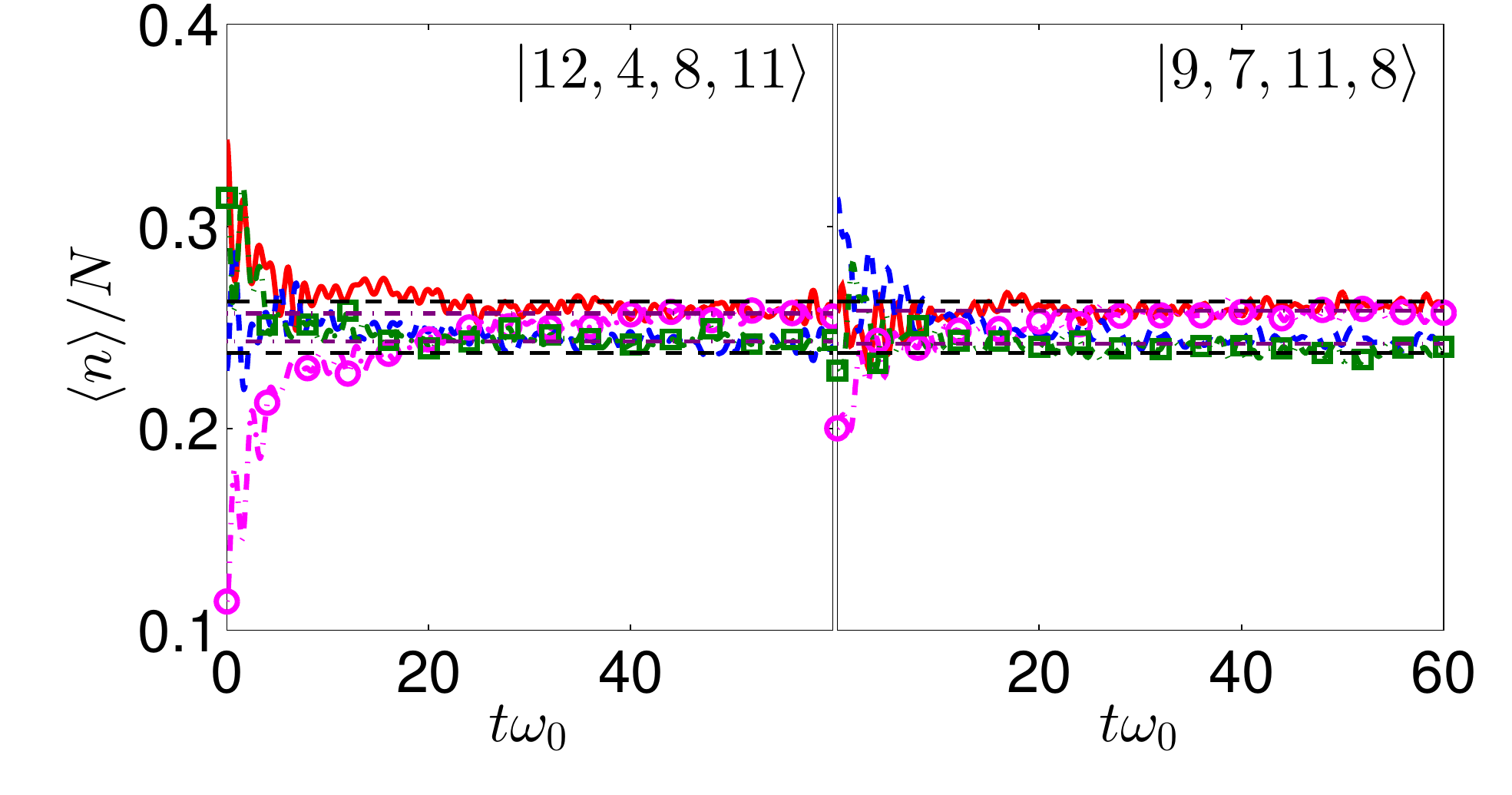}}\\
\subfloat[]{\includegraphics[width=0.85\columnwidth]{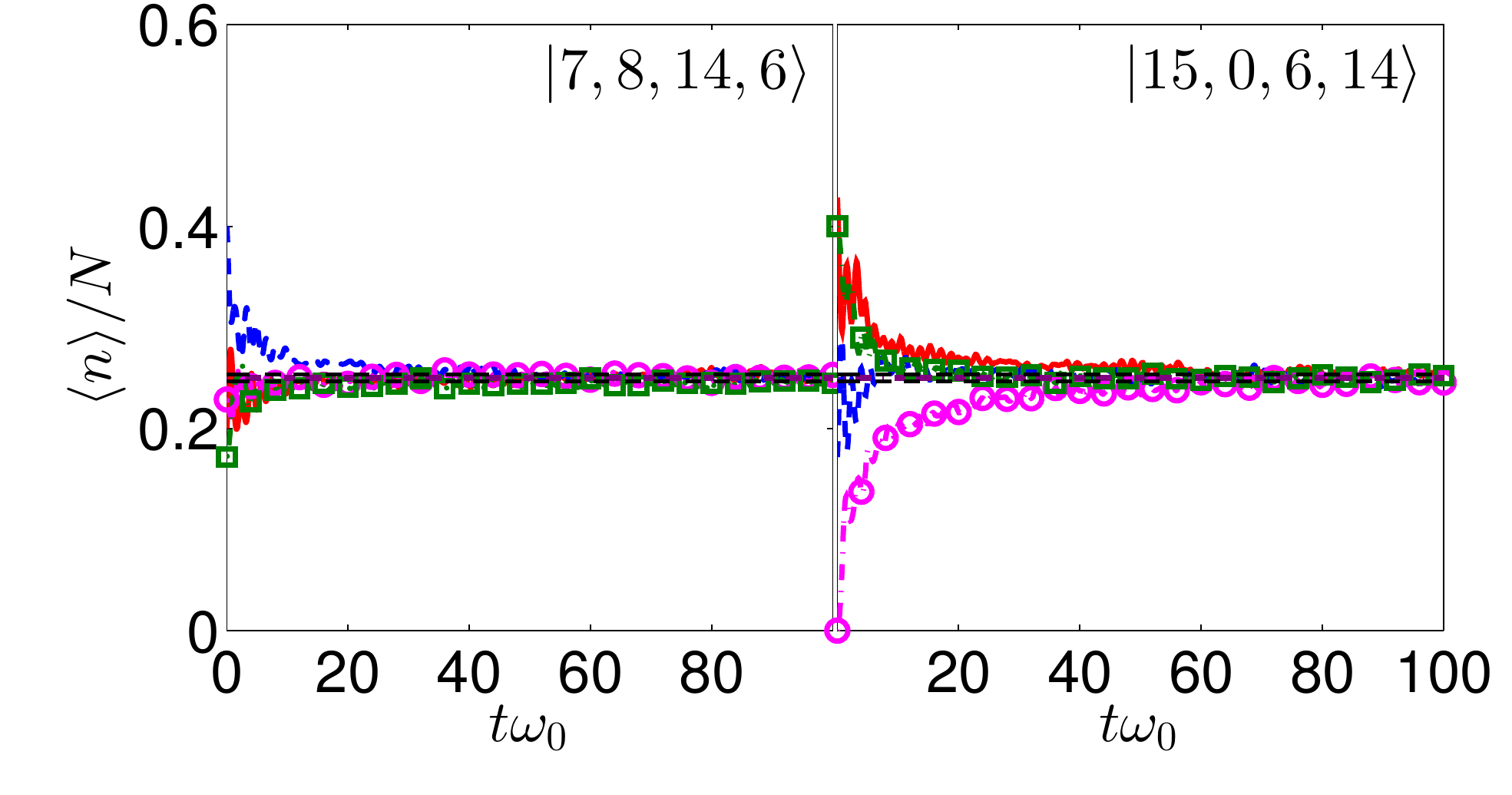}}\\
\subfloat[]{\includegraphics[width=0.85\columnwidth]{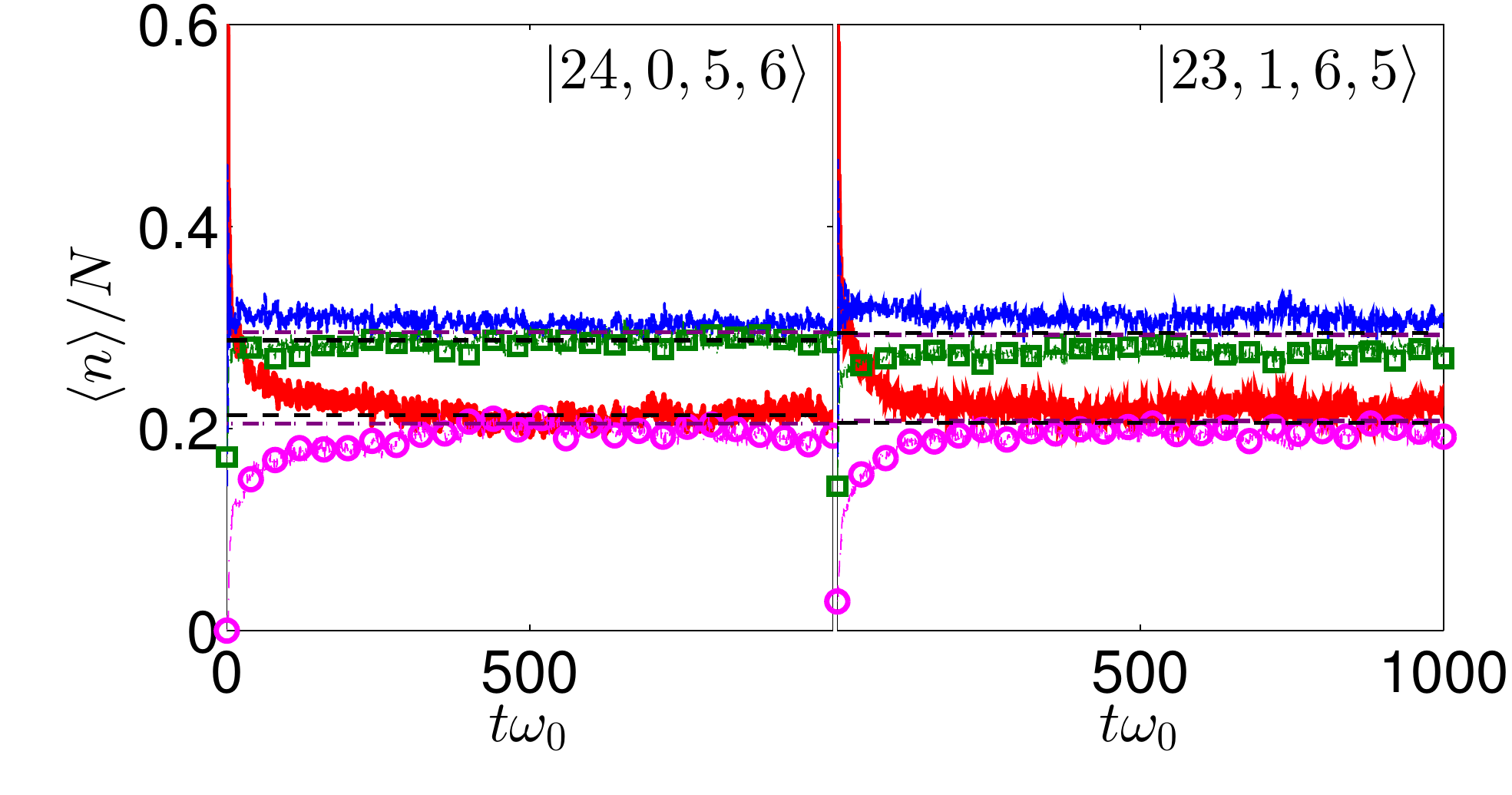}}\\
\protect\caption{(Color online) Dynamics of occupation number for Fock states with the same initial energy (from top to bottom) $E_0/\hbar\omega_0=81.053$,  $E_0/\hbar\omega_0=99.202$, $E_0/\hbar\omega_0=101.181$, and $E_0/\hbar\omega_0=115.9409$. The horizontal dashed-dotted lines correspond to the diagonal ensemble (DE) averages. The horizontal dashed lines correspond to the microcanonical ensemble (ME) averages. An initial product state is denoted as $|n^0_L, n^0_R , n^1_L, n^1_R\rangle$. }
\label{fig:u4n} 
\end{figure}

To demonstrate thermalization, or the lack thereof, we calculate the dynamics of $\langle \hat{n}^{\ell}_r \rangle$ for pairs of initial Fock states. The initial states in each pair have the same energy $E_0$. 
As mentioned before, an important signature of thermalization is the independence of thermal predictions on the details of the initial state apart from conservation laws. The conserved quantities in our system are the total number of bosons and the total energy. For our calculations, we focus on $NU^0/\hbar\omega_0=4$ and we use the same set of energies as those in Sec. \ref{sec:chaos}. The system is said to thermalize if the mode occupation numbers in each pair of initial states tend to the same stationary values and if these values are similar to a corresponding microcanonical ensemble.

In Fig. \ref{fig:u4n}(a), we present the time evolution of $\langle\hat{n}^{\ell}_r\rangle$ for initial states not too close to the ground-state energy. For this pair of initial states, the exact relaxation dynamics of the mode occupation numbers settle around their diagonal ensemble values. Moreover, the corresponding microcanonical predictions are practically indistinguishable from the stationary value of $\langle \hat{n}^{\ell}_r \rangle $. In general, we find that initial states in this region of the spectrum satisfy initial-state independence. These observations are corroborated by the smooth distribution of diagonal ensemble averages in Figs.~\ref{fig:diag2}(a) and \ref{fig:diag2}(b). 

We proceed to the characterization of the relaxation dynamics for initial states around the center of the spectrum. We find that initial states in the lower half but closer to the center of the spectrum will still exhibit thermalization. Typical results for the relaxation dynamics around this energy are shown in Fig. \ref{fig:u4n}(b). As expected for initial states in the first half of the spectrum, the long-time averaged occupation number in the lower modes are greater than those in the upper modes. This will continue to be valid as we increase the initial energy until we finally reach the middle of the spectrum where the system should now relax towards an infinite temperature state. Due to the left-right symmetry of the postquench Hamiltonian, the infinite temperature state is characterized by having equal mean occupation number between the lower modes and the upper modes, i.e., $\langle \hat{n}^0_r \rangle = \langle \hat{n}^1_r \rangle = 0.25N$. This situation is quite similar to a two-level system, with energy spacing $E_2-E_1$, coupled to an external heat bath at infinite temperature $1/T = \beta=0$ such that the ratio between the average occupation numbers is $\langle n_2 \rangle / \langle n_1 \rangle = \mathrm{exp}(- \beta (E_2-E_1)) = 1$.
We present in Fig. \ref{fig:u4n}(c) examples of time evolution towards a thermal state with almost infinite temperature. 

Even though the mode occupation numbers relax to their diagonal ensemble values in Fig. \ref{fig:u4n}(b), there is a slight deviation between the microcanonical ensemble and the stationary values. This failure of the microcanonical ensemble in describing the long-time average can be explained by the fact that the energy distributions in the middle of the spectrum are quite broad, as seen in Fig. \ref{LDoSdec}(b) and Fig. \ref{LDoSdec}(c). It is also possible that the discrepancy can be traced back from finite-size effects but further investigation is required in this direction. Nonetheless, we loosely identify the stationary state in Fig. \ref{fig:u4n}(b) as thermal state since the exact temporal evolutions of the mode occupancies still relax to the long-time average and, more importantly, this state respects the parity symmetry of the system as the bosons are evenly distributed among the wells. However, we point out some hints of intermediate relaxation in Figs.~\ref{fig:u4n}(b) and \ref{fig:u4n}(c). This is clearly seen from the time evolution of $\langle \hat{n}^0_R \rangle$ in the right panel of the same figure.

Examples of dynamics in the upper half of the spectrum are presented in Fig. \ref{fig:u4n}(d) for $E_0/\hbar\omega_0 = 115.94$. Typically, the time evolution of mode occupation numbers will have strong temporal fluctuations in this region of energy. It is also quite obvious from this plot that the exact time evolution of the mode occupation number does not approach either the diagonal or the microcanonical ensemble averages. 
Instead, the steady-state values of the occupation number of the modes in each levels,
are no longer equal $\overline{\langle \hat{n}_L^{\ell} \rangle} \neq \overline{\langle \hat{n}_R^{\ell} \rangle}$. 
The stationary states in Fig. \ref{fig:u4n}(d) break the reflection symmetry of the system and therefore we identify these equilibrium states as nonthermal. This is a concrete example of prethermalization dynamics in the system. The apparent absence of thermalization is elucidated by the behavior of the EEV shown in Fig. \ref{eev2}. Around $E_0/\hbar\omega_0 \approx 115.94$, the distribution of $\langle k | \hat{n}^{\ell}_r | k \rangle$ is no longer a smooth function of $E_k$. However, it is worth mentioning that the violation of the ETH does not guarantee the emergence of prethermalized metastable states. We further investigate this prethermalization dynamics below.

\subsection{Subsystem thermalization and prethermalization}\label{sec:sub}

In order to study subsystem thermalization, we use a mode partitioning scheme different from the usual spatial or site partitioning. For our purpose, a subsystem consists of only of one mode in the system.
We calculate the dynamics of the von Neumann entropy $S_{\mathrm{vN}}$ of a subsystem in each level. The von Neumann entropy characterizes the degree of entanglement between a mode and the rest of the system. Moreover, this choice of partitioning simplifies the numerical calculation of $S_{\mathrm{vN}}$ since the reduced density matrix of a subsystem is already diagonal in the Fock basis.
This can be seen if we write the many-body wave function in terms of the Fock basis 
\begin{align}
 |\psi(t)\rangle &= \sum_{n} c_{n}(t) |n\rangle \\ \nonumber
 &= \sum_{m=0}^N \sum_{\{p\}=N-m} c_{m,\{p\}}(t) |m\rangle \otimes |p\rangle,
\end{align}
where $|m\rangle$ is the Fock state of one mode and $|p\rangle$ is the product
state of the possible occupation number in the
other three modes. The second sum over the index $\{p\}$ is restricted by the conservation of total number of bosons in the system.
We can trace out the environment degrees of freedom $|p\rangle$
and obtain the reduced density matrix of one mode, 
\begin{align}
 \hat{\rho}_s (t) &= \sum_{m=0}^N \biggl(\sum_{\{p\}=N-m} |c_{m,\{p\}}(t)|^2\biggr) |m\rangle \langle m| \\ \nonumber
 &= \sum_{m=0}^N \lambda_m(t) |m\rangle \langle m|.
\end{align}
Hence, the von Neumann entropy of a subsystem is
\begin{equation}
 S_{\mathrm{vN}}(t) = -\mathrm{tr} [\rho \mathrm{log} \rho]= -\sum_{m=0}^N \lambda_m(t) \mathrm{log} \lambda_m(t).
\end{equation}

We are only concerned with the time evolution of initial product state and so the von Neumann entropy will always start at zero. Following the quench, the von Neumann entropy is expected to increase as the time evolution couples the modes with one another.
We now study the dynamics of the von Neumann entropy for the same set of parameters and initial states used in Sec. \ref{sec:relmod}. If the $S_{\mathrm{vN}}$ were to equilibrate, it should saturate at a value predicted by a Gibbs ensemble. Since the number of bosons and the energy are both not conserved in the subsystem, the appropriate Gibbs ensemble must be the grand canonical one. The number of bosons and the energy in a subsystem are obtained using the local operators
\begin{equation}
	\hat{n}_s = \hat{n}^{\ell}_r;~\hat{H}_s = U^{\ell}\hat{n}^{\ell}_r(\hat{n}^{\ell}_r-1) +E^{\ell}_r \hat{n}^{\ell}_r,
\end{equation}
respectively. Therefore, the density matrix of the grand canonical ensemble is
\begin{equation}
 \hat{\rho}_{\mathrm{GC}} = \frac{1}{Z}e^{-\beta(\hat{H}_s - \mu \hat{n}_s)} |m\rangle \langle m|,
\end{equation}
where $Z=\mathrm{tr}(\hat{\rho}_{\mathrm{GC}})$.  
The inverse temperature $\beta$ and
the chemical potential $\mu$ are fixed by the conditions (i) $ \overline{\langle \hat{H}_s \rangle} = \mathrm{tr} (\hat{H}_s \hat{\rho}_{\mathrm{GC}} )$ 
and (ii)$ \overline{\langle \hat{n}_s \rangle} = \mathrm{tr} (\hat{n}_s \hat{\rho}_{\mathrm{GC}} )$, where the overline denotes the expectation value obtained from the diagonal ensemble. To calculate $ \overline{\langle \hat{H}_s \rangle}$, we need the long-time average of the two-body local operator $\langle (\hat{n}^{\ell}_r)^2 \rangle$ obtained in Sec. \ref{sec:deme}.

\begin{figure}[!ht]
\subfloat[]{\includegraphics[width=1\columnwidth]{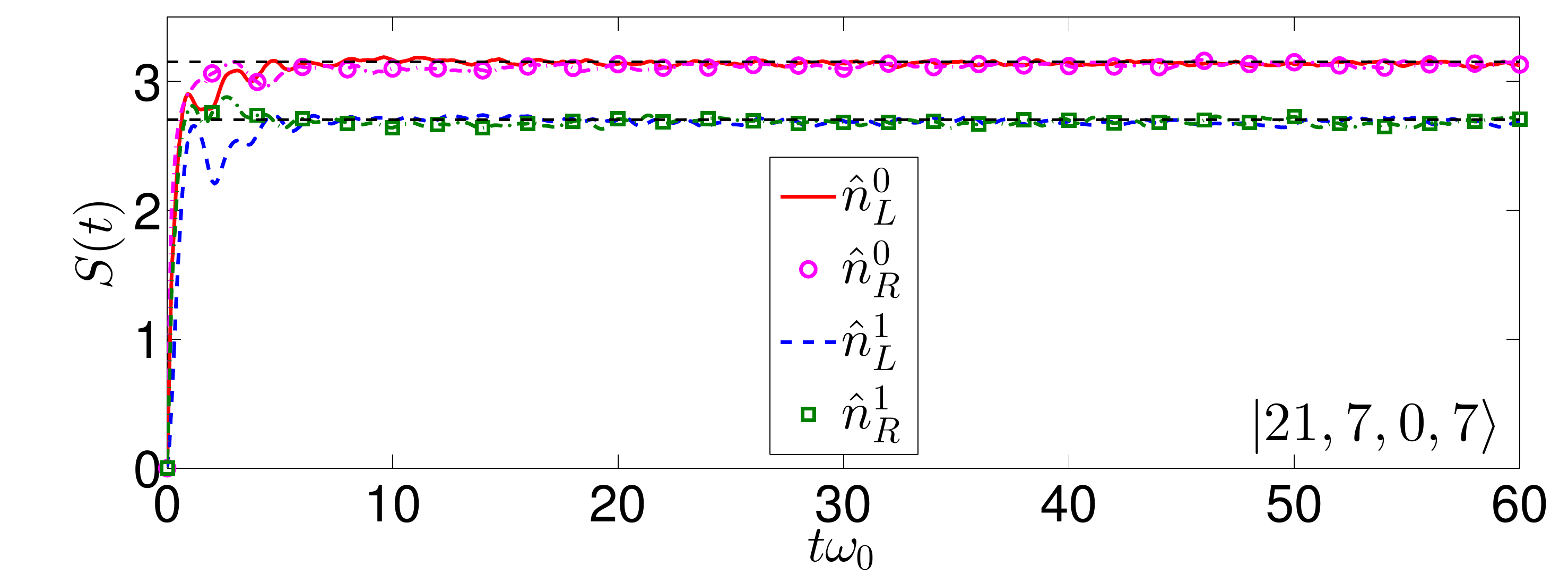}}\\
\subfloat[]{\includegraphics[width=1\columnwidth]{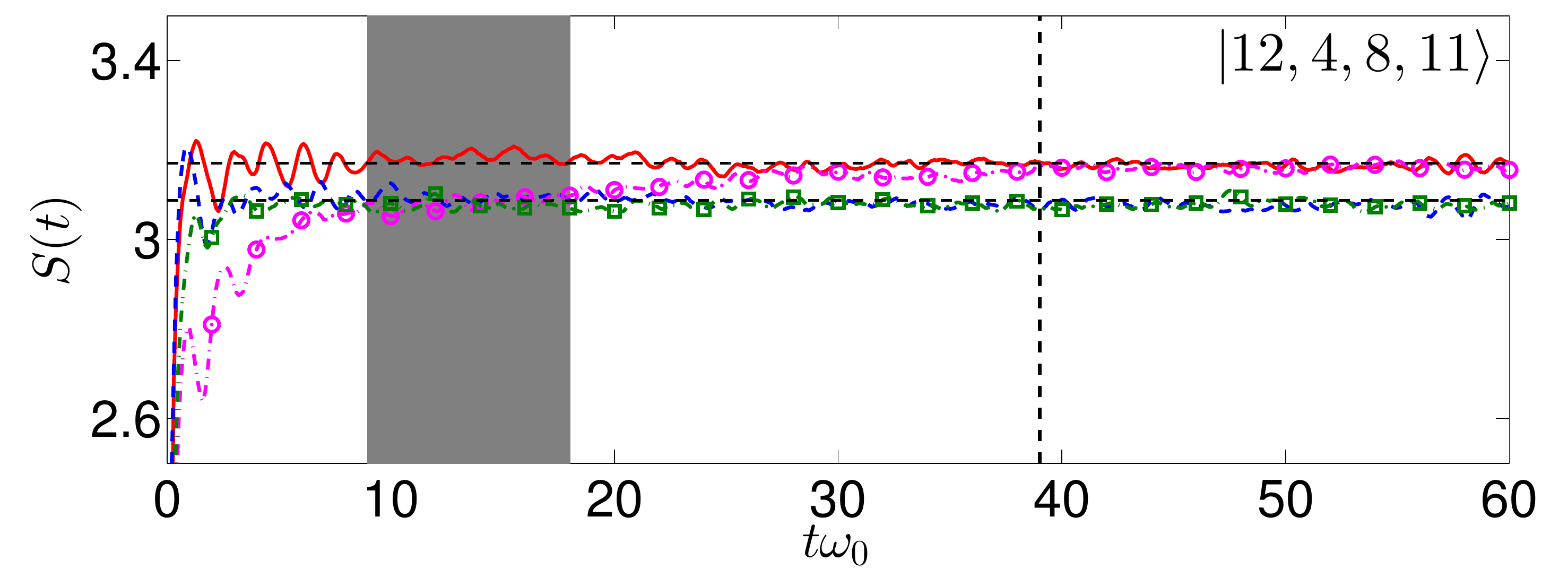}}\\
\subfloat[]{\includegraphics[width=1\columnwidth]{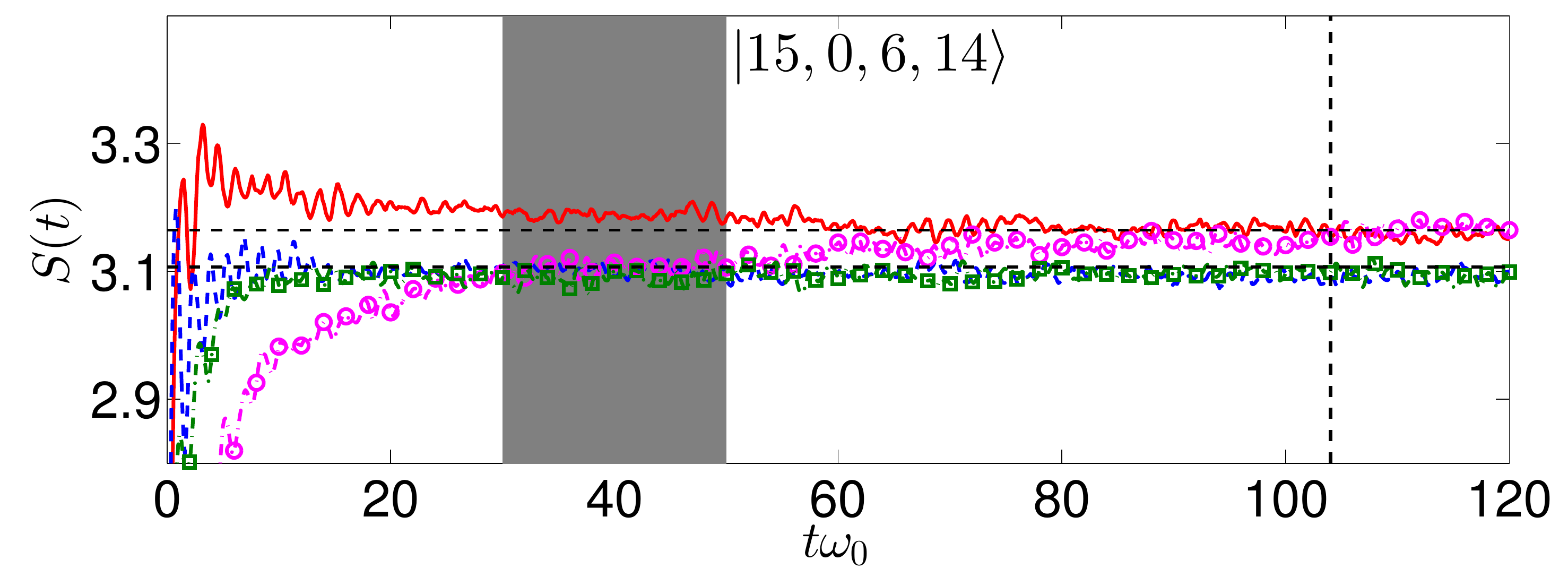}}\\
\subfloat[]{\includegraphics[width=1\columnwidth]{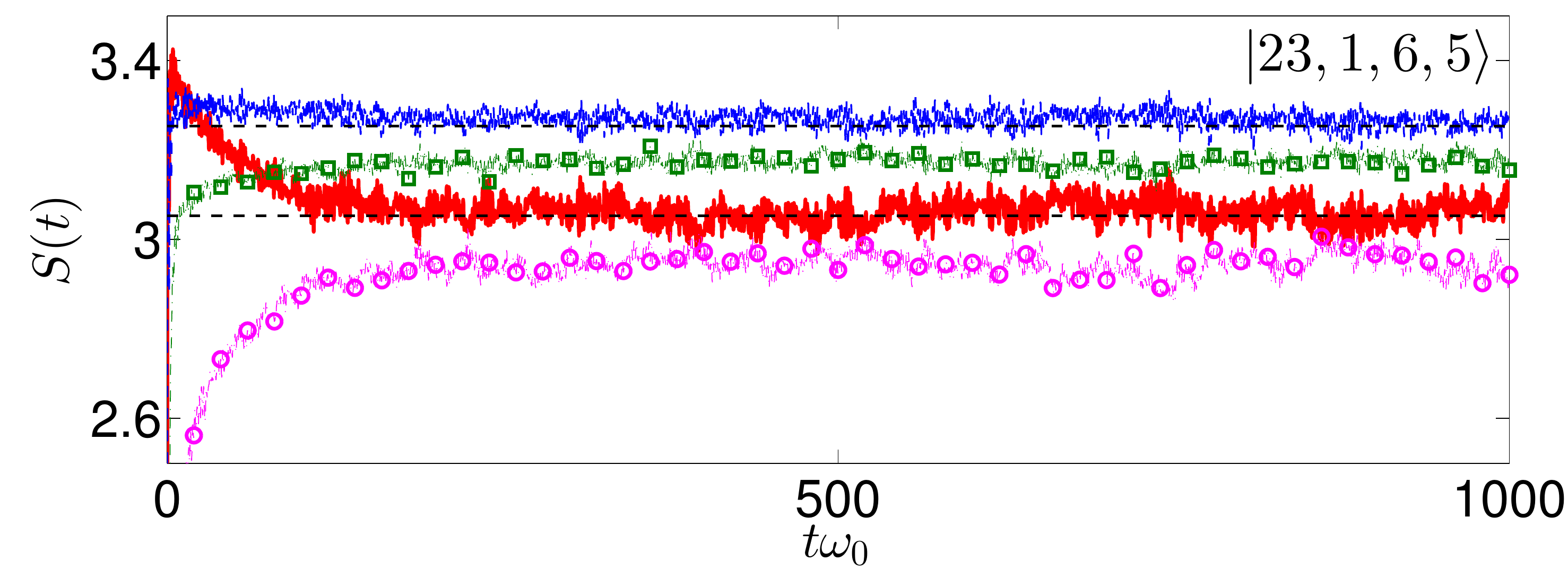}}
\caption{(Color online) Time evolution of the von Neumann entropy compared to grand canonical entropy (horizontal dashed lines) for (a) $E_0/\hbar\omega_0=81.05$, (b) $E_0/\hbar\omega_0=99.20$, (c)  $E_0/\hbar\omega_0=101.18$, and (d)  $E_0/\hbar\omega_0=115.94$. The shaded area emphasizes the prethermalization plateaus. The vertical dashed lines correspond to the thermalization time after the prethermalized regimes in panels (b) and (c). }
\label{vNe} 
\end{figure}

Typical time evolution of the von Neumann entropy is shown in Fig. \ref{vNe}. We focus on the long-time dynamics and the general features of relaxation towards the steady-state values (or quasi-steady-state values).
After the rapid growth in the von Neumann entropy, the behavior of the ensuing dynamics largely depends on the energy of the initial product state. For initial energy near the ground-state energy, the von Neumann entropy quickly approaches the Gibbs entropy in a single-step relaxation.  Furthermore, the thermalization time is roughly equal to that of the mode occupation number. This is exemplified in Fig. \ref{vNe}(a).

For initial states still in the lower half but closer to the middle of the spectrum, we observe a two-step relaxation process such as those presented in Figs.~\ref{vNe}(b) and \ref{vNe}(c). 
The von Neumann entropy will first relax to a quasistationary value different from the thermal entropy. 
The metastable states manifest as prethermalization plateaus in the dynamics of the von Neumann entropy.
We remark that not all of the modes will exhibit such prethermalization dynamics.  
For example, in Fig. \ref{vNe}(b), the modes in the upper levels of both wells are already fluctuating close to the grand canonical entropy while the mode in the lower level of the right well is still trapped in a prethermalized state. Similar behavior is seen in Fig. \ref{vNe}(c), but here the lifetime of the prethermalized state is longer. 
The lifetime of the prethermalization plateau for $E_0/\hbar\omega_0=99.20$ is $t\omega_0 \sim 10$ while for $E_0/\hbar\omega_0=101.18$ it is $t\omega_0 \sim 20$. In general, we find that the time scale of the prethermalized regimes increases as the initial energy of the system increases. Consequently, the thermalization time in Fig. \ref{vNe}(b) ($t\omega_0 \sim 39$) is smaller than that in Fig. \ref{vNe}(c) ($t\omega_0 \sim 103$).
The prethermalized states in this part of the spectrum can be physically interpreted as metastable states that temporarily break the parity symmetry of the postquench Hamiltonian. The plateaus observed in the dynamics of the von Neumann entropy reveal the disagreement between the prethermalized entropy of a mode in one well and its counterpart on the other well. The symmetry is later restored as the system evolves towards parity-conserving thermal states. 

Initial states found in the upper half of the spectrum, such as those in Fig. \ref{vNe}(d), appear to get trapped in extremely long-lived prethermalized states, at least within the numerically accessible time scales. In  Fig. \ref{vNe}(d), the von Neumann entropies of each mode in the left well have already reached the Gibbs prediction but the modes in the right well still fluctuate around a nonthermal value.
It is worth noting that the metastable states found in this energy region do not show clear signs of any drift towards thermal equilibrium. 
This behavior seemingly contradicts known results for prethermalization in other nearly integrable models (see Refs. \cite{Kollar2011,Bertini2015}) but there is no reason to rule out with certainty the possibility of such metastable states to decay after sufficient time. This issue is left for future work.

\section{Prethermalization and Time scales}\label{sec:pretherm}
In this section, we use similar arguments as in Ref.~\cite{Gong2013} to understand the emergence of prethermalized states in the system.
We split the expression for the time evolution of a generic local operator into three contributions,
\begin{align}\label{atherm}
\langle \hat{A} \rangle &=\sum_{k \neq l,\{k,l\} \notin \{a,b\}}C^{k*}_{n_0} C^{l}_{n_0}  A_{lk} e^{i(E_l-E_k)t/\hbar}   \\ \nonumber
	& + \sum_{a \neq b}C^{a*}_{n_0} C^{b}_{n_0} A_{ba} e^{i\delta_{ab}t/\hbar} +\sum_k |C^k_{n_0}|^2 A_{kk}, 
\end{align}
where $\{a,b\}$ is a set of indices corresponding to the pairs of quasidegenerate eigenstates with energy difference $\{\delta_{ab}=E_b-E_a\}$. 
Written this way, it is possible to associate two different timescales due to the first two sums in Eq. \eqref{atherm}. 
For brevity, we refer to each sum in Eq. \eqref{atherm}, in order of its appearance, as the nondegenerate, the quasidegenerate, and the diagonal contributions. The nondegenerate contribution is associated with a thermalization timescale due to dephasing, as shown in Eq.~\eqref{ethex}.
It is worth mentioning that the ETH already puts a strong restriction on the off-diagonal elements, $A_{kk'}=\langle k | \hat{A} |k' \rangle$, being small relative to its diagonal counterpart. Instead, we turn our attention to possible interplay between the structure of the spectrum and the coefficients $C^{k}_{n_0}$ .

The energy distribution of an initial state plays an important role in understanding the stages of relaxation dynamics in our system as they provide crucial information on the chaoticity or sparsity of the coefficients $C^k_{n_0}$. 
The width of the energy shell measures the connectivity of an initial state to the eigenstates of the final Hamiltonian.
An initial state that thermalizes will never be fully extended with respect to the eigenstates as seen in Fig. \ref{LDoSdec} and therefore the width of the LDoS will limit contributions in Eq. \eqref{atherm} of eigenstates with energies far from the initial energy. This naturally justifies a separation of the spectrum into different sectors. We can then analyze the structure of the eigenenergies within each sector. 

\begin{figure}[!ht]
\begin{center}
{\includegraphics[width=0.5\columnwidth]{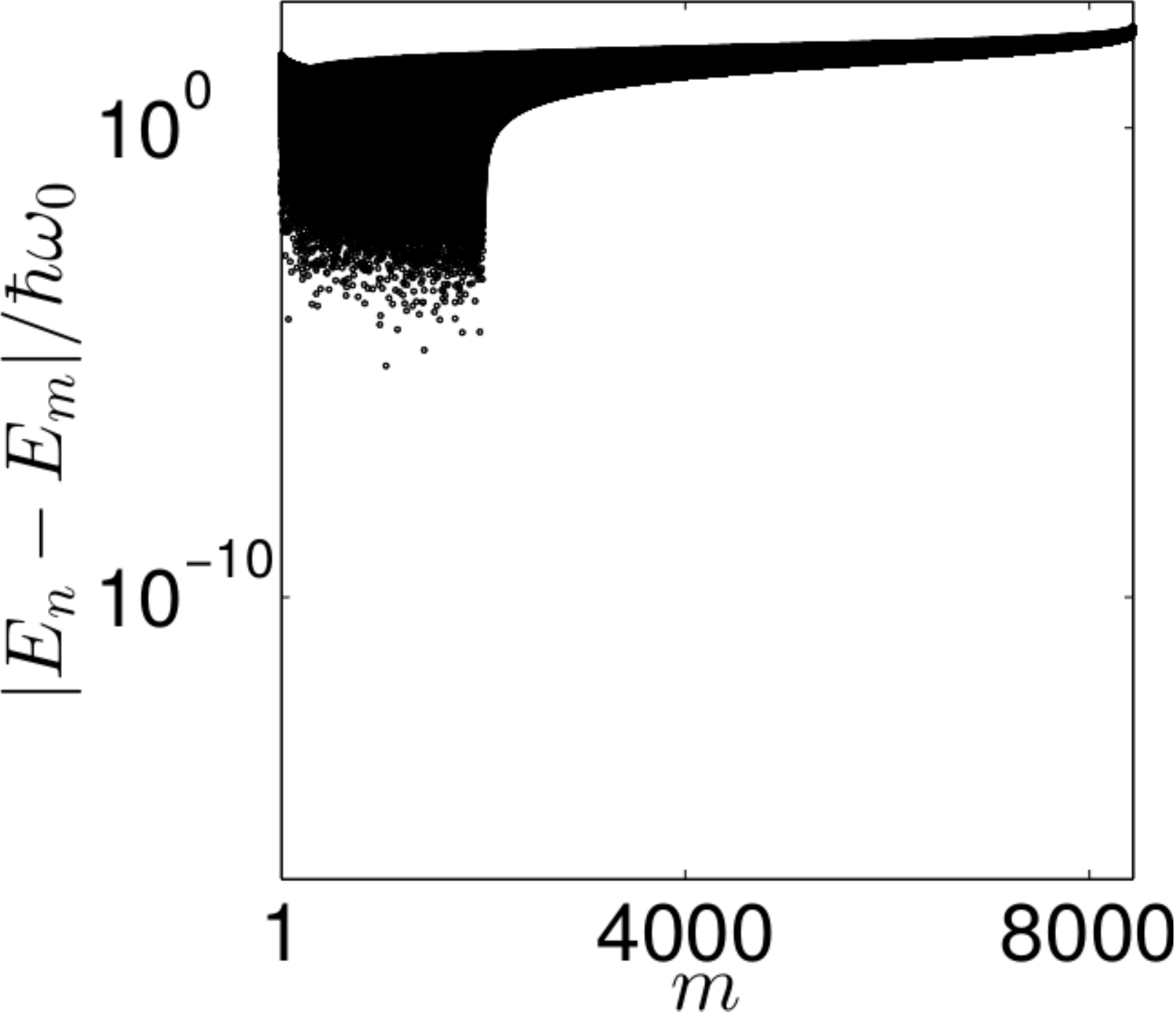}}{\includegraphics[width=0.5\columnwidth]{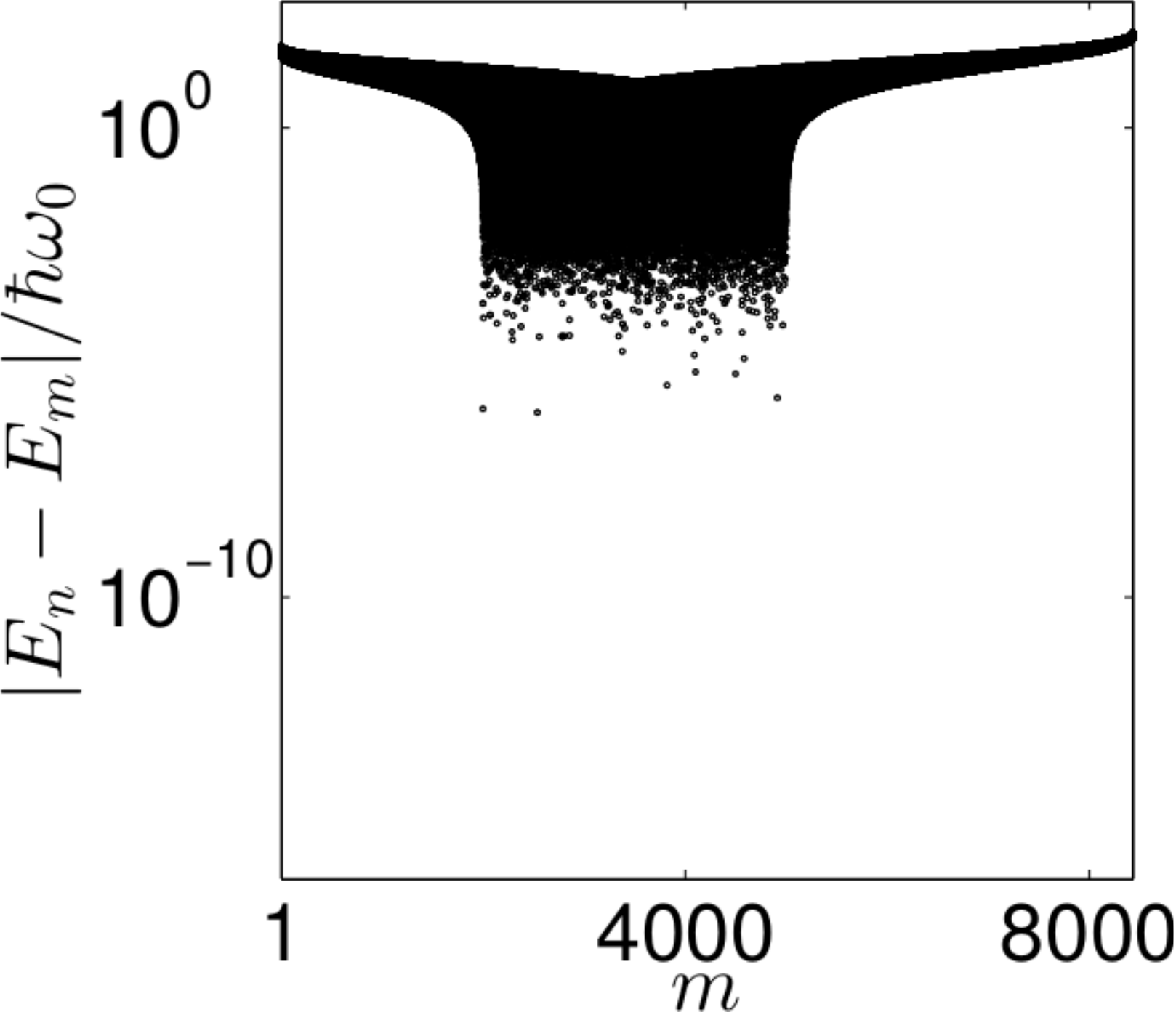}}
\par\bigskip
{\includegraphics[width=0.5\columnwidth]{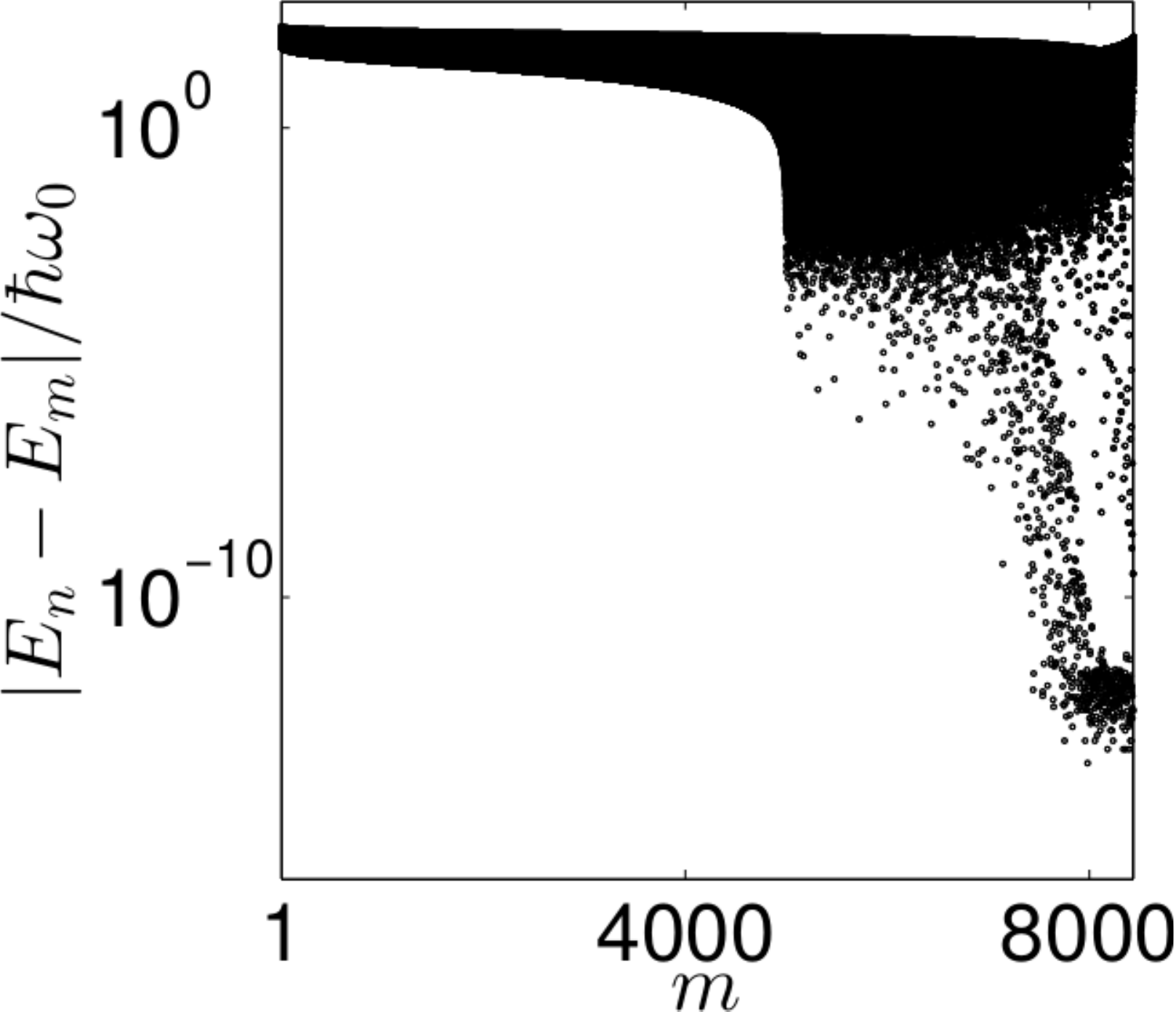}}
\protect\caption{(Color online) Energy differences across the spectrum. The eigenenergies $E_n$ are arranged in increasing order $n \in [1,D]$, where $D$ is the dimension of the Hilbert space. (Top left) Low-energy sector $n \in [1,2000]$. (Top right) Middle-energy sector $n \in [2001,5000]$. (Bottom) High-energy sector $n \in [5001,D]$.}
\label{enem}
\end{center} 
\end{figure}

We divide the full spectrum into three sectors in accordance to the choice of initial energies in Secs. \ref{sec:chaos} and \ref{sec:relmod}; i.e., $E_0/\hbar\omega_0=81.05$ is in the low-energy sector; both $E_0/\hbar\omega_0=\{99.20,~101.18\}$ are in the middle-energy sector; and $E_0/\hbar\omega_0=115.94$ is in the high-energy sector. 
The results are presented in Fig. \ref{enem}. The outliers in each plots represent quasidegeneracies of pairs of even and odd eigenstates.
The energy differences, in general, decrease from $\sim 10^{-5}$ in the low-energy sector to $\sim 10^{-6}$  in the middle sector. This explains the increase in the observed thermalization time from $t \lesssim 10^1$ for initial states with low energy to $t \sim 10^2$ for initial energies in the midspectrum.
Also, there is perceivable increase in the prevalence of quasidegenerate pairs as you go to higher energies.
This means that for sufficiently high initial energies, the quasidegenerate contribution in Eq. \eqref{atherm} becomes more dominant and an intermediate time scale may emerge. Thus, we conjecture that the presence of quasidegeneracies led to the prethermalization observed in the system. Moreover, the width of the energy shell restricts the connectivity among the $C^k_{n_0}$ such that only the eigenenergies near an initial energy will contribute significantly to the relaxation dynamics. For this reason, initial states in the low-energy sector are unaffected by the quasidegeneracies in the high-energy sector and we only find fast and single-stage relaxation for such initial states [see Figs.~\ref{fig:u4n}(a) and \ref{vNe}(a)]. In contrast, metastable states start to appear during the relatively slower relaxation dynamics of initial states in the middle sector [see Figs.~\ref{fig:u4n}(b), \ref{fig:u4n}(c), \ref{vNe}(b), and \ref{vNe}(c)]. 
One of the main results of this work is this crossover behavior from single-stage to two-stage relaxation process without changing the parameters of the Hamiltonian. Instead, we can probe this transition just by increasing the energy of the initial product state.
On the other hand, most of the quasidegeneracies are situated in the high-energy sector. In addition to this, the energy difference can reach as low as $\sim 10^{-15}$ in this sector, which is several orders of magnitude lower than the other sectors. These two factors combined may explain why initial states in the high-energy sector exhibit long-lived prethermalization plateaus [see Fig. \ref{fig:u4n}(d) and \ref{vNe}(d))].

\section{Summary and Conclusion}\label{sec:conc}

In this work, we have numerically investigated the relaxation dynamics following an integrability-breaking quench in a double-well system. 
We obtained the distribution of consecutive level spacings in order to characterize the spectral statistics of the system. Then, we identified the postquench Hamiltonian as nonchaotic over a wide range of interaction parameters due to the absence of level repulsion. The enhanced level clustering for strong interactions is attributed to the increase in the amount of quasidegenerate pairs. Thus, for the chosen trap parameters, the nonintegrable postquench Hamiltonian is close to an integrable point.

In order to check the requirements for the validity of the ETH in our system, we obtained the distribution of the EEV for various interaction parameters. In general, smooth distributions of the EEV are found in the lower half of the spectrum. By changing the interaction strength, we observed that the distribution of the EEV broadens as the number of quasidegenerate energy levels increases. 
At the moment, we only provided numerical evidence that the ETH holds for the system, albeit having a spectral statistics akin to integrable models. 
It would be interesting to see whether the weak ETH scenario studied in Refs. \cite{Biroli2010,Alba2015} can be observed in our system. This can be a subject of future study involving careful scaling analysis of the model. 

We compared the long-time averages of local operators calculated using the diagonal ensemble and their microcanonial values for a set of initial product states spanning all possible combinations of Fock states in each mode. The range of energy where the diagonal ensemble averages are well described by the microcanonial ensemble is consistent with that for which the ETH is satisfied. Therefore, we have numerically verified that it is possible for the system to thermalize.
We computed the LDoS of typical initial states with energy within the first half of the energy spectrum and we classified them as chaotic since they ergodically fill the energy shell. These results allowed us to extend the scope of the main conclusion in Ref. \cite{Torres2013} to include chaotic initial states away from the middle of the spectrum that thermalize under time evolution of a nonchaotic Hamiltonian.

Using generic initial product states, we found that the system may exhibit thermalization in the sense that the exact time evolution of local operators relax to the diagonal ensemble predictions and these values are close to corresponding Gibbs ensemble predictions. In particular, we demonstrated that the mode occupation numbers of certain initial states would dynamically approach the diagonal ensemble averages. The energy of such initial state is within the range of energy for which initial states are delocalized and the ETH is valid.
Observables for initial states in the high-energy region are shown to relax but not towards the diagonal ensemble values. This led us to think about the possibility of having two-stage relaxation dynamics in the system. In fact, the emergence of prethermalized states became clear when we examined the time evolution of the von Neumann entropy in each mode. Specifically, we have observed intermediate relaxation to metastable states for initial energies close to the middle of the spectrum. Finally, we have argued that delocalization of initial states together with the details of the energy spectrum such as quasidegeneracies play vital roles in understanding prethermalization and the relaxation time scales of the system. One of the key contributions of this work is the possibility of probing, by simply changing initial product states, the continuous transition from single-step to two-step relaxation at fixed Hamiltonian parameters.

One possible direction of future work is a careful study on how finite-size effects may affect the phenomenon observed in this work including the emergence of prethermalization
and the behavior of the EEV as the size of the system is increased. Also, the question of whether an initial state with LDoS different from the energy shell but still has some kind of structure, as in Fig. \ref{LDoSdec}(d), will eventually thermalize remains an open issue for future study.

\acknowledgments
We acknowledge O. Fialko for introducing the author to the topic of this work. We also thank J. Brand, Y.~Y. Atas, and S. Flach for suggestions and useful discussions.

%\bibliography{ref}
%

\end{document}